\documentclass[fullpaper,final]{nldl}

\paperID{7}
\vol{352}

\usepackage[utf8]{inputenc}
\usepackage[T1]{fontenc}
\usepackage{amsmath}
\usepackage{amsfonts}
\usepackage{mathtools}
\usepackage{nicefrac}
\usepackage{booktabs}
\usepackage{multirow}
\usepackage{graphicx}
\usepackage{textcomp}
\usepackage{float}
\usepackage{enumitem}

\usepackage{algorithm}
\usepackage{algorithmic}

\usepackage{hyperref}
\usepackage{url}
\hypersetup{
  pdfusetitle,
  colorlinks,
  linkcolor = BrickRed,
  citecolor = NavyBlue,
  urlcolor  = Magenta!80!black,
}
\usepackage[capitalize,nameinlink]{cleveref}

\graphicspath{{figures/}}

\setlist[description]{nosep,leftmargin=2.5cm,labelwidth=2.3cm,labelsep=0.2cm,
  font=\normalfont\bfseries}

\usepackage[acronym,nogroupskip,nopostdot,nonumberlist]{glossaries}
\makeglossaries
\glsdisablehyper

\newacronym{dm}{DM}{diffusion model}
\newacronym{ddpm}{DDPM}{denoising diffusion probabilistic model}
\newacronym{ddim}{DDIM}{denoising diffusion implicit model}
\newacronym{ldm}{LDM}{latent diffusion model}
\newacronym{gan}{GAN}{generative adversarial network}
\newacronym{cgan}{cGAN}{conditional generative adversarial network}
\newacronym{cnn}{CNN}{convolutional neural network}
\newacronym{mt}{MT}{microtubule}
\newacronym{irm}{IRM}{interference reflection microscopy}
\newacronym{tirf}{TIRF}{total internal reflection fluorescence}
\newacronym{ct}{CT}{computed tomography}
\newacronym{mri}{MRI}{magnetic resonance imaging}
\newacronym{fid}{FID}{Fr\'echet inception distance}
\newacronym{kid}{KID}{kernel inception distance}
\newacronym{dfid}{DINO-FID}{Fr\'echet inception distance computed in DINOv2 feature space}
\newacronym{dkid}{DINO-KID}{kernel inception distance computed in DINOv2 feature space}
\newacronym{emd}{EMD}{Earth Mover's distance}
\newacronym{iou}{IoU}{intersection-over-union}
\newacronym{skiou}{SKIoU}{skeletonised intersection-over-union}
\newacronym{kl}{KL}{Kullback--Leibler}
\newacronym{sam}{SAM}{Segment Anything Model}
\newacronym{ema}{EMA}{exponential moving average}
\newacronym{cielab}{CIELAB}{CIE $L^*a^*b^*$ colour space}
\newacronym{spade}{SPADE}{spatially-adaptive denormalization}
\newacronym{snr}{SNR}{signal-to-noise ratio}
\newacronym{pet}{PET}{positron emission tomography}

\definecolor{collapseRed}{HTML}{D55E00}   
\definecolor{healthyGreen}{HTML}{009E73}  

\newcommand{\PlaceholderFig}[3][0.55\textheight]{%
  \IfFileExists{figures/#2}{%
    \includegraphics[width=\linewidth]{#2}%
  }{%
    \fbox{\parbox[c][#1][c]{0.96\linewidth}{\centering%
      \textsc{\small Placeholder figure}\\[0.5em]%
      \texttt{figures/#2}\\[0.4em]%
      {\itshape\small #3}}}%
  }%
}

\newcommand{\afcNExperts}{11}
\newcommand{\afcNTrials}{110}
\newcommand{\afcNCorrect}{56}
\newcommand{\afcAcc}{50.9\%}
\newcommand{\afcCIlo}{41.7\%}
\newcommand{\afcCIhi}{60.1\%}
\newcommand{\afcBinomP}{0.92}
\newcommand{\afcEquivDelta}{10}

\newcommand{\afcEquivHi}{60\%}
\newcommand{\afcTostP}{0.033}
\newcommand{\afcEquivMargin}{8.6}
\newcommand{\afcNReal}{33}
\newcommand{\afcNFake}{33}
\newcommand{\afcNAnon}{146}
\newcommand{\afcPooledN}{256}

\newcommand{\afcPooledAcc}{47.7\%}
\newcommand{\afcPooledCIlo}{41.6\%}
\newcommand{\afcPooledCIhi}{53.8\%}
\newcommand{\afcPooledBinomP}{0.49}
\newcommand{\afcLeftRate}{46.4\%}
\newcommand{\afcLeftAccR}{0.15}
\newcommand{\afcMedianRT}{4.3}
\newcommand{\afcMedianRTcorrect}{3.9}
\newcommand{\afcMedianRTincorrect}{5.2}
\newcommand{\afcMinExpertAcc}{20.0\%}
\newcommand{\afcMaxExpertAcc}{80.0\%}

\providecommand{\tstrReal}{\tstrPLACE}
\providecommand{\tstrVone}{\tstrPLACE}
\providecommand{\tstrVtwo}{\tstrPLACE}
\providecommand{\tstrVlast}{\tstrPLACE}
\providecommand{\tstrRealVtwo}{\tstrPLACE}

\providecommand{\skiouHuman}{0.80}
\providecommand{\skiouHumanStd}{0.14}
\providecommand{\tstrNFolds}{4}

\providecommand{\hpoNSeeds}{10}
\providecommand{\hpoNTrials}{200}
\providecommand{\hpoModels}{SAM3Text, MicroSAM and CellSAM}
\providecommand{\hpoVoneMedLo}{\tstrPLACE}
\providecommand{\hpoVoneMedHi}{\tstrPLACE}
\providecommand{\hpoVtwoMedLo}{\tstrPLACE}
\providecommand{\hpoVtwoMedHi}{\tstrPLACE}
\providecommand{\hpoSamVone}{\tstrPLACE}
\providecommand{\hpoSamVtwo}{\tstrPLACE}
\providecommand{\hpoSamVoneLo}{\tstrPLACE}
\providecommand{\hpoSamVoneHi}{\tstrPLACE}

\renewcommand{\tstrReal}{0.66}
\renewcommand{\tstrVone}{0.43}
\renewcommand{\tstrVlast}{0.52}
\renewcommand{\tstrVtwo}{0.67}
\renewcommand{\tstrRealVtwo}{\textbf{0.69}}

\renewcommand{\hpoVoneMedLo}{0.45}
\renewcommand{\hpoVoneMedHi}{0.66}
\renewcommand{\hpoVtwoMedLo}{0.51}
\renewcommand{\hpoVtwoMedHi}{0.71}
\renewcommand{\hpoSamVone}{0.66}
\renewcommand{\hpoSamVoneLo}{0.47}
\renewcommand{\hpoSamVoneHi}{0.76}
\renewcommand{\hpoSamVtwo}{0.71}

\newcommand{\synmtone}{\textit{SynthMT}\xspace}
\newcommand{\synmtv}{\textit{DiffuMT}\xspace}

\newcommand{\runFifteenEpoch}{315}          
\newcommand{\runFifteenDiversityEpoch}{250} 

\title{Diagnosing Diversity Collapse and Validating Mask-Conditioned Diffusion for Labeled Microtubule Microscopy}

\author[1]{Mario Koddenbrock}
\author[1]{Frederic Rapp}
\author[2,3]{Simone Reber}
\author[1]{Erik Rodner}
\affil[1]{HTW Berlin}
\affil[2]{Berliner Hochschule für Technik (BHT), Berlin, Germany}
\affil[3]{Max Planck Institute for Infection Biology, Berlin, Germany}
\affil[ ]{\texttt{\{mario.koddenbrock, erik.rodner\}@htw-berlin.de}}

\begin{document}

\maketitle

\begin{abstract}
Mask-conditioned diffusion models have become standard for generating labeled training data in biomedical imaging. But the tools used to evaluate them were built for a different problem. Standard metrics like \gls{fid} and \gls{kid} score images through ImageNet-trained features. Those features do not transfer to microscopy, leaving the metrics poorly calibrated for specialized domains. Even with a domain-appropriate feature space, a single blended number can still hide a real failure: a model can produce visually plausible images that still transfer poorly to downstream tasks.
We identify a specific failure mode: Late in training, realism keeps improving while texture diversity collapses. The model settles onto a fixed color palette, converging to a high-similarity, low-diversity state that standard metrics do not penalize.
Therefore, we propose a diversity-aware diagnostic in a domain-validated feature space. It combines a realism axis based on inter-similarity to real images with a diversity axis based on intra-similarity among generated samples.
This diagnostic selects a checkpoint that domain experts cannot reliably distinguish from real recordings in a forced-choice study.
It also yields useful downstream segmentation: a segmenter trained only on synthetically labeled data reaches a median \gls{skiou} comparable to one trained on real data, at lower cross-fold variance, and clearly ahead of the best available public alternative in this domain, a parametric renderer.
We also reproduce a data-efficient hyperparameter-transfer experiment from prior work, tuning several foundation segmenters on a small labeled subset and evaluating on real images: transfer is stronger with our data.
We release \synmtv on HuggingFace, including the triplet dataset, code to reproduce the downstream-utility validation, and a standalone diagnostic tool for evaluating mask-conditioned diffusion models.
\end{abstract}

\section{Introduction}
\label{sec:intro}

\Glspl{mt} are cytoskeletal filaments that drive chromosome segregation, intracellular transport, and cell motility~\citep{garcin2019microtubules}.
Their dynamic instability is the target of widely used chemotherapeutics, including taxanes such as paclitaxel~\citep{prota2023structural}, as well as emerging neuroprotective therapies for neurodegeneration~\citep{soliman2022microtubule}.
To quantify how candidate drugs perturb microtubule dynamics, \glspl{mt} are reconstituted \emph{in~vitro} from stabilized nucleation seeds and imaged using \gls{irm} or \gls{tirf} microscopy~\citep{koddenbrock2026synthetic}.
Extracting quantitative measurements such as filament counts, lengths, and curvature distributions requires accurate instance segmentation.
However, pixel-precise tracing of overlapping thin filaments remains a labor-intensive task, with expert annotation taking minutes per image.
Recorded microscopy data therefore quickly exceeds available labeling capacity.

This challenge is not unique to microtubule microscopy.
Across biomedical imaging, including colonoscopy~\citep{dorjsembe2024polypddpm}, anatomical \gls{ct} and \gls{mri}~\citep{konz2024,nazir2025diffaug}, pixel-level expert annotations remain a major bottleneck.
A common solution is to generate additional labeled data using \emph{mask-conditioned diffusion models}~\citep{konz2024,dorjsembe2024polypddpm,nazir2025diffaug,liu2025conditional,wang2025diffusionprism}.
A binary annotation specifies where a structure should appear, while the diffusion model generates a corresponding realistic image.
Pixel-space mask conditioning~\citep{konz2024,dorjsembe2024polypddpm} preserves fine spatial details that latent-space approaches such as ControlNet~\citep{zhang2023controlnet} may attenuate.
However, training these models on small domain-specific datasets can still result in severe mode collapse (\cref{sec:experiments}).

The difficulty has therefore shifted from generating synthetic biomedical images to evaluating them.
The standard metrics for generative image quality evaluation, including \gls{fid}~\citep{heusel2017fid} and \gls{kid}~\citep{binkowski2018kid}, were developed for natural-image benchmarks with large and diverse datasets.
Recent studies have shown that these metrics are poorly calibrated when transferred to small biomedical domains~\citep{stein2023exposingflaws,ravuri2023understanding}.
Furthermore, mask conditioning introduces an additional challenge: when the mask provides most of the spatial information, generated textures may collapse while image-level similarity remains high.
A model can therefore appear realistic according to standard metrics while producing synthetic data that is insufficiently diverse or ineffective for downstream analysis.

\begin{figure*}[tb]
\centering
\includegraphics[width=\textwidth]{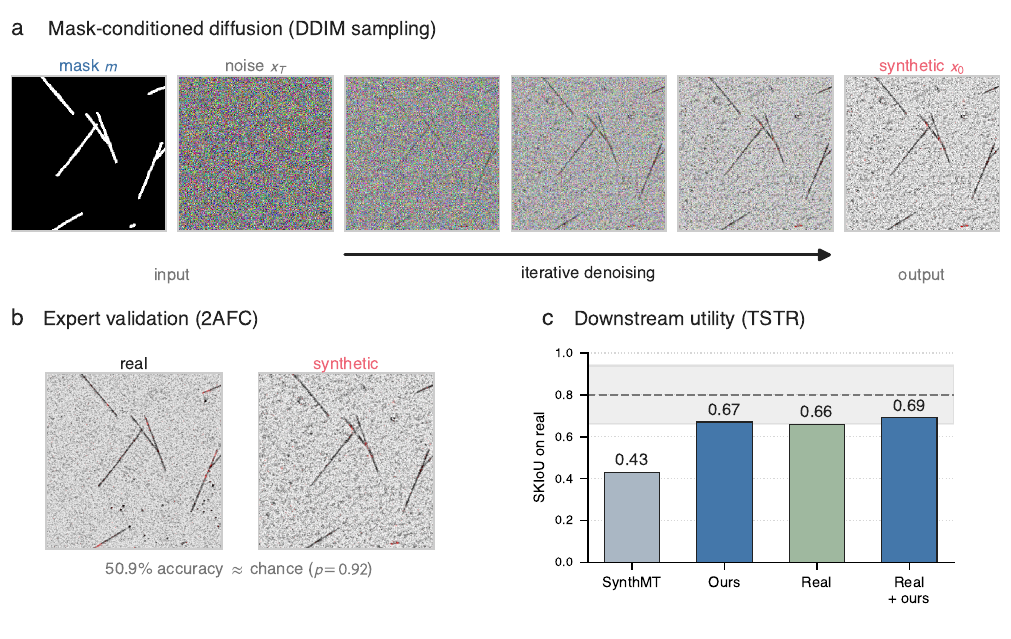}
\caption{\textbf{The diagnostic-selected checkpoint is both perceptually indistinguishable from real and useful for downstream segmentation.}
\textbf{(a)}~\emph{Generation:} a binary mask~$m$ and Gaussian noise~$x_T$ are denoised by a mask-conditioned DDPM (DDIM sampling) into a synthetic image~$x_0$, selected by our diagnostic rather than FID; the denoising chain is illustrative.
\textbf{(b)}~\emph{Expert validation:} shown a real and a generated image of the same mask, experts pick the real one only at chance: they cannot reliably tell synthetic from real.
\textbf{(c)}~\emph{Downstream utility:} a segmenter trained only on synthetic data and tested on real (train-synthetic/test-real) matches a real-trained model and far exceeds the parametric baseline; the dashed line is the inter-annotator agreement on real images (the achievable segmentation ceiling).
Exact values, uncertainties and protocols are reported in \cref{sec:twoafc,sec:tstr}.}
\label{fig:overview}
\end{figure*}

We address this problem by evaluating mask-conditioned diffusion models through both realism and downstream utility (\cref{fig:overview}).
We train a pixel-space mask-conditioned \gls{dm} on \gls{irm} microtubule recordings and evaluate generated images using a two-alternative-forced-choice (2AFC) study with domain experts.
Experts cannot reliably distinguish generated images from real recordings, demonstrating high perceptual realism.

Realism alone, however, is not sufficient.
Synthetic data must also support accurate analysis of real microscopy images.
We therefore evaluate synthetic-data-only training for microtubule instance segmentation and reproduce a data-efficient hyperparameter-transfer experiment from prior work.
A segmenter trained only on our synthetic data approaches the performance of one trained on real data, substantially outperforming the strongest available public parametric baseline.
Furthermore, repeating a prior hyperparameter-transfer study shows that our synthetic data improves foundation-segmenter adaptation to real \gls{irm} images.

\Cref{fig:teaser} highlights the failure mode motivating our diagnostic.
Across checkpoints of the same mask-conditioned diffusion model, realism measured against real images remains nearly unchanged, while diversity varies substantially.
A checkpoint selected by realism alone can therefore correspond to a collapsed generation regime.
We propose a diversity-aware diagnostic in a domain-validated feature space (DINOv2), which jointly measures a realism axis (inter-similarity to real images) and a diversity axis (intra-similarity within the generated set).
The diagnostic identifies checkpoints that balance realism and diversity and selects the model that passes the expert 2AFC evaluation.

\paragraph{What this paper contributes.}
\begin{itemize}\itemsep0.15em
\item \textbf{\synmtv, validated by 2AFC.}
We release an open triplet dataset consisting of binary masks, real \gls{mt} images, and their mask-conditioned synthetic counterparts.
Domain experts cannot reliably distinguish our generated images from real recordings in a 2AFC study (\cref{sec:twoafc}).
\footnote{Live 2AFC challenge: \iffinal\url{https://huggingface.co/spaces/HTW-KI-Werkstatt/DiffuMT}\else \textit{[URL withheld for double-blind review.]}\fi}

\item \textbf{A diversity-aware diagnostic for mask-conditioned diffusion.}
We show that realism metrics alone fail to detect late-stage diversity collapse and introduce a diagnostic that jointly evaluates realism and diversity in a domain-validated feature space (\cref{sec:method,sec:exp-degeneracy}).

\item \textbf{Proof of downstream utility.}
A segmenter trained only on \synmtv achieves performance on real microtubule images comparable to real-data training while substantially outperforming the strongest public synthetic baseline.
We further show improved adaptation of foundation segmenters using synthetic data in a reproduced low-data tuning experiment (\cref{sec:tstr,sec:hpo-transfer}).
\end{itemize}
\section{Related work}
\label{sec:related}

\paragraph{Diffusion models.}
Denoising diffusion models~\citep{sohldickstein2015deep,ho2020ddpm} and the related score-based formulation~\citep{song2019score} are the de-facto generators for high-resolution image synthesis~\citep{dhariwal2021beatgans,nichol2021iddpm}.
Sampling is accelerated by \gls{ddim}~\citep{song2021ddim}.
It is made tractable at high resolution by latent diffusion~\citep{rombach2022ldm}.
It is most commonly conditioned via classifier-free guidance~\citep{ho2022cfg}.

\paragraph{Fidelity and diversity in generative evaluation.}
Distributional metrics such as \gls{fid}~\citep{heusel2017fid} and \gls{kid}~\citep{binkowski2018kid} combine sample quality and diversity into a single score.
Even within their original natural-image setting, Inception-based scores do not consistently track human judgments of image quality~\citep{nguyen2024generating}.
This has motivated metrics that explicitly separate these factors, including precision and recall~\citep{kynkaanniemi2019improved}, density and coverage~\citep{naeem2020reliable}, and the reference-free Vendi score~\citep{friedman2023vendi}.
Our diagnostic follows this decomposition: inter-similarity to real images captures realism, while intra-similarity within generated samples captures diversity.
Unlike previous metrics, we study their behavior in mask-conditioned biomedical diffusion, where realism often saturates while diversity collapses.
The two axes must therefore be interpreted jointly and calibrated against a domain-specific reference distribution rather than ImageNet-based defaults~\citep{kynkaanniemi2023role,stein2023exposingflaws}.

\paragraph{Conditioned diffusion in biomedical imaging.}
Diffusion models have recently been adopted for biomedical image synthesis and analysis~\citep{kazerouni2023diffusionmedical}.
Conditioning can be performed in latent feature spaces, as in ControlNet~\citep{zhang2023controlnet} and \gls{spade}~\citep{park2019spade}, or directly through spatial mask concatenation.
The latter preserves pixel-level control and has been applied to synthetic medical images, including breast \gls{mri}, abdominal \gls{ct}, and colonoscopy~\citep{konz2024,dorjsembe2024polypddpm}.
For microtubule microscopy, previous diffusion approaches focused on unconditional super-resolution~\citep{saguy2023microtubule}, whereas our goal is mask-conditioned generation of labeled \gls{irm} data.
However, mask-conditioned diffusion can still collapse when conditioning provides most of the spatial information.
Recent methods such as Diffusion Prism~\citep{wang2025diffusionprism} and CADS~\citep{sadat2024cads} address this through training or sampling modifications; we instead focus on diagnosing and selecting against this failure mode.

\paragraph{Offset noise and zero-terminal-\gls{snr}.}
Standard diffusion schedules leave residual signal at the final timestep, creating a train--inference mismatch that can bias samples toward the dataset mean.
Offset noise~\citep{guttenberg2023offset} and related zero-terminal-\gls{snr} formulations~\citep{lin2024commondnoise} address this issue by enabling broader luminance variation.
We adopt offset noise (\cref{sec:method}) to avoid low-contrast attractors in \gls{irm} microscopy.

\paragraph{2AFC validation in biomedical generators.}
Two-alternative forced choice (2AFC) is a standard psychophysical protocol for measuring whether observers can distinguish generated from real images.
It has been used to evaluate biomedical generators, including medical diffusion models~\citep{hung2023medcdiff} and synthetic oncology imaging~\citep{fedrigo2023tgan}.
We use 2AFC as a domain-level validation of perceptual realism, complementing metric-based evaluation with downstream utility.

\section{Method}
\label{sec:method}

We evaluate a mask-conditioned diffusion model (\cref{sec:model}) using a diversity-aware diagnostic (\cref{sec:diag}).
The diffusion model itself follows \citet{konz2024}; our contribution is the evaluation framework.

\subsection{The model under study}
\label{sec:model}

We use a \gls{ddpm}~\citep{ho2020ddpm,sohldickstein2015deep} sampled with \gls{ddim}~\citep{song2021ddim} and conditioned on binary segmentation masks in pixel space following \citet{konz2024}.
The denoiser $\epsilon_\theta$ is a 2D U-Net~\citep{ronneberger2015unet} that receives the channel-wise concatenation of the noisy image $x_t \in \mathbb{R}^{3 \times h \times w}$ and mask $m \in \{0,1\}^{h \times w}$:
\begin{equation}
\mathcal{L}_m =
\mathbb{E}_{x_0,m,t,\epsilon}
\left[
\|\epsilon-\epsilon_\theta(x_t,t|m)\|^2
\right],
\end{equation}
where
$x_t=\sqrt{\bar{\alpha}_t}x_0+\sqrt{1-\bar{\alpha}_t}\epsilon$.
Although DDIM sampling is deterministic for a fixed initial noise, different noise realizations produce different valid images, defining a conditional distribution $p(x|m)$.
Therefore, each released triplet represents only one possible synthetic realization of a mask.
The only practical modifications to \citet{konz2024} (\cref{app:impl}) is offset noise~\citep{guttenberg2023offset}.

\subsection{A three-axis diagnostic}
\label{sec:diag}

The diagnostic evaluates three complementary properties: realism, diversity, and appearance statistics.
No individual axis reliably detects collapse; their joint behavior distinguishes healthy and degenerate regimes.

\paragraph{Plausibility and diversity (DINOv2).}
We embed images using the fifth block of DINOv2~\citep{oquab2024dinov2}, following the domain validation of \citet{koddenbrock2026synthetic} and consistent with evidence that intermediate-layer features can be more transferable than network outputs~\citep{bolya2025perception}, and normalize embeddings with $L_2$ normalization.
Realism is measured by \emph{inter-similarity}, the average cosine similarity between generated images and their $k=10$ nearest real neighbors.
Diversity is measured by \emph{intra-similarity}:
\begin{align}
S_\mathrm{intra} &=
\frac{1}{n(n-1)}
\sum_{i\neq j}
\hat{y}_i^\top\hat{y}_j ,
\end{align}
where lower values indicate higher diversity.
This formulation places realism and diversity on the same feature scale.
It correlates with the Vendi score~\citep{friedman2023vendi} across all ablations ($|r|=0.98$, \cref{app:vendi}).
We check this dependence on DINOv2 by substituting four other embedding models (\cref{app:model-ablation}) and by varying the DINOv2 layer (\cref{app:layer-ablation}).
We additionally report DINOv2-based Fréchet and kernel distances (\gls{dfid}, \gls{dkid}) and a memorization delta~\citep{carlini2023extracting,somepalli2023diffusion}.
These metrics are reported for completeness but are not used for checkpoint selection.

\paragraph{Mask fidelity.}
Because generated images have no ground-truth segmentation, we evaluate mask consistency by re-segmenting generated images with SAM~3~\citep{sam32025}.
We compare the recovered masks to the conditioning masks using \gls{skiou}~\citep{liu2018skiou} and F1 at thresholds $0.5$ and $0.75$.
Scores are interpreted relative to human agreement (\gls{skiou}$=\skiouHuman$).

\paragraph{Color and brightness.}
Images are converted to \gls{cielab} space and separated into foreground and background regions using the conditioning mask.
For each region, we measure the \gls{emd} between generated and reference distributions.
This captures appearance collapse that may not be visible from structural metrics alone.

\paragraph{Aggregation.}
For checkpoint selection, we combine the three perceptual axes (Inter, Intra, Color) using a geometric mean:
\begin{equation}
S_\text{global}
=
\left(
\prod_{a\in\mathcal{A}}\hat{s}_a
\right)^{1/|\mathcal{A}|},
\label{eq:agg}
\end{equation}
where all axes are min-max normalized with higher values indicating better quality.
The geometric mean prevents one strong axis from compensating for a collapsed one.
But the diagnostic is robust to the aggregation choice; arithmetic and geometric means select the same checkpoint (\cref{app:agg-robustness}).
\section{Experiments}
\label{sec:experiments}

This section first establishes the failure mode motivating our diagnostic (\cref{sec:exp-degeneracy}) before validating the selected checkpoint through expert 2AFC evaluation (\cref{sec:twoafc}). Experimental details are given in \cref{sec:setup}.

\subsection{Experimental Setup}
\label{sec:setup}

\paragraph{Data.}
The dataset was originally curated by \citet{koddenbrock2026synthetic} and comes from \gls{irm} recordings acquired in two laboratories with different microscope systems and tubulin sources.
\glspl{mt} are thin, low-contrast filaments visible against a tightly controlled \emph{in vitro} background.
We extract random $512\times512$ crops, 309 of which are manually annotated across 75 distinct source recordings.
These are re-tiled into $256\times256$ patches with 128-pixel overlap for training (no overlap for testing), yielding 2961 training tiles, 200 annotated test tiles, and 1336 unlabeled reference patches used for similarity scoring.

\paragraph{Training.}
All runs share the same 113M-parameter 2D U-Net backbone, optimizer, batch size, and \gls{ddim} sampler; full implementation details are given in \cref{app:impl}.
We use offset noise with scale $\sigma=0.2$, and vary only the $\beta$ schedule (linear vs.\ scaled cosine) and regression loss (MSE, $L_1$, Huber) in the ablation study.
The selected checkpoint is robust to the offset-noise scale (\cref{app:sigma-robustness}); additional implementation details and ablations are deferred to the appendix.

\subsection{Inter-similarity alone is insufficient}
\label{sec:exp-degeneracy}

\begin{figure}[tb]
\centering
\setlength{\fboxsep}{0pt}
\setlength{\fboxrule}{1.5pt}
\begin{subfigure}[t]{0.31\linewidth}
  \centering
  \fcolorbox{collapseRed}{white}{\includegraphics[width=\linewidth]{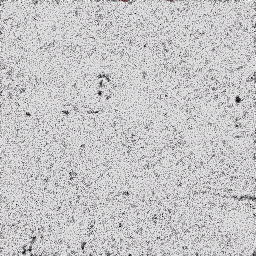}}\\[0.25em]
  {\scriptsize Inter $0.98$\\ Intra {\color{collapseRed}\textbf{$0.99$}}}
\end{subfigure}\hfill
\begin{subfigure}[t]{0.31\linewidth}
  \centering
  \fcolorbox{healthyGreen}{white}{\includegraphics[width=\linewidth]{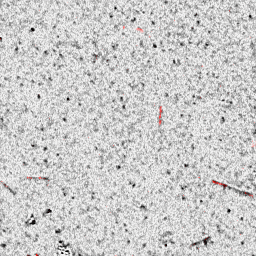}}\\[0.25em]
  {\scriptsize Inter $0.97$\\ Intra {\color{healthyGreen}\textbf{$0.87$}}}
\end{subfigure}\hfill
\begin{subfigure}[t]{0.31\linewidth}
  \centering
  \fcolorbox{collapseRed}{white}{\includegraphics[width=\linewidth]{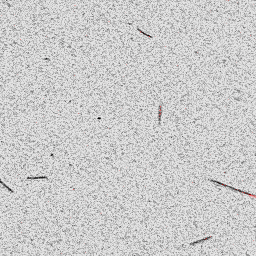}}\\[0.25em]
  {\scriptsize Inter $0.98$\\ Intra {\color{collapseRed}\textbf{$0.97$}}}
\end{subfigure}
\caption{\textbf{Inter-similarity gives only a weak, small-magnitude signal that does not reliably separate collapsed from healthy ($0.98/0.97/0.98$).}
Three checkpoints of \emph{one} run, conditioned on the \emph{same}
binary mask, shown left to right: an early checkpoint (collapsed, blurry; no valid \glspl{mt} are visible),
a middle one (realistic and diverse), and a late one (collapsed to a
near-uniform texture).
Intra-similarity flags the two collapsed checkpoints (red), and only the
middle checkpoint matches the real diversity baseline (green).}
\label{fig:teaser}
\end{figure}

\begin{table*}[tb]
\centering
\caption{
Comparison of our diagnostic-selected model (\synmtv), the parametric baseline (\synmtone), and the naïve final checkpoint.
Although the naïve checkpoint achieves the best \gls{dfid}, it exhibits substantially lower diversity and worse color fidelity, illustrating why \gls{dfid} alone is insufficient for checkpoint selection.
The real column reports inter-annotator agreement for mask fidelity and reference statistics for all other metrics.
}
\label{tab:comparison}
\small
\begin{tabular}{lcccc}
\toprule
\textbf{Metric} & \textbf{Real} & \synmtone & \synmtv & \textbf{Naive} \\
                & \textit{(ground truth)} & \textit{(parametric,~prior~work)} & \textit{(ours)} & \textit{(no~selection)} \\ \midrule
\gls{skiou} ($\uparrow$) & $0.80 \pm 0.14$ & $0.96 \pm 0.15$ & $0.86 \pm 0.23$ & $0.86 \pm 0.10$ \\
F1@50 ($\uparrow$)              & $0.89 \pm 0.10$ & $0.98 \pm 0.15$ & $0.93 \pm 0.26$ & $0.99 \pm 0.10$ \\
F1@75 ($\uparrow$)              & $0.87 \pm 0.11$ & $0.97 \pm 0.16$ & $0.88 \pm 0.33$ & $0.90 \pm 0.30$ \\
\midrule
\acrshort{dfid} ($\downarrow$)              & -- & $0.27$ & $0.14$ & $\mathbf{0.07}$ \\
\midrule
Color, global ($\downarrow$)              & -- & $8.93$ & $\mathbf{3.56}$ & $3.86$ \\
Color, foreground ($\downarrow$)          & -- & $12.87$ & $\mathbf{4.92}$ & $6.62$ \\
Color, background ($\downarrow$)          & -- & $9.19$ & $2.70$ & $\mathbf{2.45}$ \\
\midrule
Inter-Sim. ($\uparrow$)         & -- & $0.86 \pm 0.01$ & $\mathbf{0.97 \pm 0.01}$ & $0.96 \pm 0.01$ \\
Intra-Sim. ($\approx 0.88$)         & $0.88 \pm 0.02$ & $0.94 \pm 0.03$ & $\mathbf{0.89 \pm 0.03}$ & $0.93$ \\
$\Delta$ Intra-Sim. ($\downarrow$)  & -- & $0.06$ & $\mathbf{0.01}$ & $0.05$ \\
\bottomrule
\end{tabular}%
\end{table*}

\paragraph{Inter-similarity saturates while diversity collapses.}

\Cref{fig:teaser} illustrates the central failure mode.
Across three representative checkpoints, inter-similarity changes little despite clear visual differences, whereas intra-similarity correctly identifies the healthy checkpoint.
The same pattern persists throughout training (\cref{fig:epoch-trajectories}): inter-similarity rises rapidly before plateauing, while intra-similarity first improves and then collapses.
Late in training, the two measures become positively correlated ($r=+0.58$), indicating increasing realism at the expense of diversity.
Color statistics corroborate the same trend ($r=-0.71$, $p<0.001$), whereas \gls{dfid} and \gls{dkid} remain largely insensitive, demonstrating that distributional realism alone does not reveal late-stage diversity collapse.
The selected checkpoint does not minimize intra-similarity alone, but achieves the best balance across all diagnostic axes.
This behavior is consistent across all eight hyperparameter ablations and directly motivates our diversity-aware checkpoint selection.
A qualitative counterpart is shown in \cref{fig:diversity-grid}, where repeated sampling from the same conditioning mask produces diverse yet plausible textures at the selected checkpoint.

\paragraph{Numerical signature.}
\Cref{fig:binned-corr} makes this precise by binning checkpoints into three training stages and correlating metrics within each bin.
Inter- and intra-similarity are decoupled early and in the middle of training ($r\approx+0.09$), so neither is a proxy for the other; they only re-couple in the late bin, which is the collapse itself rather than a mere correlate of it.
The Vendi score~\citep{friedman2023vendi} tracks intra-similarity almost perfectly in every bin ($|r|\geq0.97$), confirming that our diversity axis is not an idiosyncratic choice (\cref{app:vendi}).
The full per-bin correlation structure, including the color and \gls{dfid} axes, is given in \cref{app:binned-corr}.

\paragraph{Comparison with existing approaches.}

\Cref{tab:comparison} compares our diagnostic-selected model (\synmtv), the parametric baseline \synmtone~\citep{koddenbrock2026synthetic}, and a naïve last-epoch checkpoint.
The naïve model highlights the limitation of conventional checkpoint selection: despite matching \synmtv's mask fidelity and achieving the best \gls{dfid}, it exhibits substantially lower diversity and poorer color fidelity.
Conversely, \synmtv achieves the best balance across realism, diversity, and color fidelity, demonstrating that our diagnostic avoids selecting visually collapsed checkpoints.
The parametric baseline reproduces conditioning masks almost perfectly but deviates substantially in appearance statistics, illustrating why mask fidelity alone is insufficient for evaluating generative models.
Additional comparisons across hyperparameter configurations are provided in \cref{app:ablation}.

\section{Expert validation (2AFC)}
\label{sec:twoafc}

The diagnostic of \cref{sec:diag} selects a single best checkpoint across all hyperparameter configurations and training epochs (offset~$0.2$, linear $\beta$-schedule, MSE loss, epoch~\runFifteenEpoch{}).
We thank a reviewer for pointing out that our checkpoint-selection reference pool included training images; recomputing the selection with a corrected, held-out reference pool ranks epoch~\runFifteenEpoch{} second of 200 checkpoints (score $0.7433$, vs.\ $0.7855$ for the new top-ranked epoch). this gap may reflect sampling noise rather than a genuine quality difference (\cref{app:corrected-reference}).
We retain epoch~\runFifteenEpoch{} as the model validated below.
To quantify the realism of the resulting model independently of any metric used to select it, we run a two-alternative-forced-choice (2AFC) study with \afcNExperts{} domain experts on \afcNReal{} real and \afcNFake{} generated $256\!\times\!256$ images drawn from the held-out test set.
The participant pool combines two groups of domain experts: cell-biology and microscopy researchers who routinely work with \gls{irm} recordings of \gls{mt} dynamics, and machine-learning researchers who routinely inspect generative-model outputs.
None of them was involved in selecting, training, or curating the model.

\begin{figure}[tb]
\centering
\includegraphics[width=\linewidth]{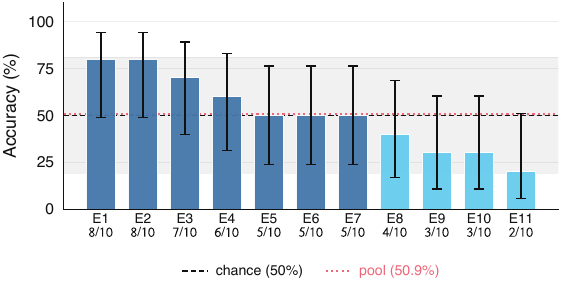}
\caption{\textbf{Experts cannot distinguish generated from real images: pool aggregate $p=\afcBinomP{}$.}
One bar per expert; $95\%$ Wilson CIs; dashed line $=$ chance, dotted line $=$ pool mean (\afcAcc{}).}
\label{fig:2afc-per-expert}
\end{figure}

\paragraph{Protocol.}
Each trial presents two images side by side, one real and one generated, drawn at random from the pool and randomly swapped left/right, and asks the expert to identify the real one.
The generated images are conditioned on the masks of the corresponding real images, so a real/generated pair shares the same underlying filament geometry (\cref{fig:overview}).
The trials are administered through the public HuggingFace Space that also serves as an open community challenge.\footnote{Live 2AFC challenge: \iffinal\url{https://huggingface.co/spaces/HTW-KI-Werkstatt/DiffuMT}\else \textit{[URL withheld for double-blind review.]}\fi}
The full protocol is given in \cref{app:2afc}.

\paragraph{Result.}
Experts could not reliably distinguish synthetic from real images: aggregate accuracy was \afcAcc{} (\afcNCorrect{}/\afcNTrials{} trials; $95\%$ Wilson CI $[\afcCIlo{},\afcCIhi{}]$), statistically consistent with chance ($p=\afcBinomP{}$, exact binomial test).
A two-one-sided equivalence test further confirms that performance is within $\pm\afcEquivMargin{}$ percentage points of chance ($p=\afcTostP{}$), supporting perceptual indistinguishability rather than simply failing to detect a difference.
Per-expert results are consistent with the binomial null (\cref{fig:2afc-per-expert}); however, inference is performed at the pool level since each expert contributed only $10$ trials.
The equivalence-margin choice, two protocol controls (positional bias and engagement), and a robustness check on the anonymous trials are detailed in \cref{app:2afc}.

\paragraph{Comparison to the prior expert study on \synmtone.}
The parametric predecessor \synmtone~\citep{koddenbrock2026synthetic} was perceptually evaluated through a $7$-point Likert protocol on five aspects (structural fidelity, background, lighting, noise, overall) with $n=6$ experts; the authors chose Likert ratings over pairwise forced-choice to obtain aspect-specific judgments.
Their reported outcome is a ``measurable gap to real images [that] persists across all dimensions''.
Our chance-level 2AFC on \synmtv closes that gap on the strictest test available.

\section{Downstream utility and transfer}
\label{sec:downstream}

Perceptual realism alone does not guarantee useful synthetic data.
We therefore evaluate whether \synmtv improves downstream segmentation of real \gls{mt} images in two settings: training a segmenter on synthetic data (\cref{sec:tstr}) and transferring tuned inference parameters from synthetic to real (\cref{sec:hpo-transfer}).
Both use \gls{skiou} as the evaluation metric and are interpreted relative to human agreement on real images (\skiouHuman$\pm$\skiouHumanStd) reported in \citep{koddenbrock2026synthetic}.

\subsection{Train on synthetic, test on real}
\label{sec:tstr}

\begin{figure}[tb]
\centering
\includegraphics[width=\linewidth]{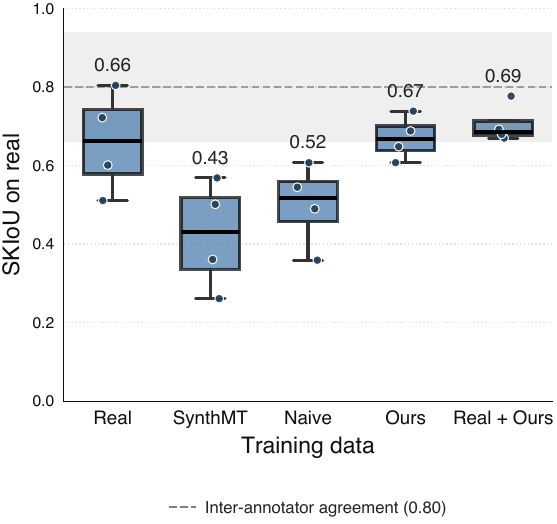}
\caption{\textbf{The diagnostic-selected synthetic data is useful downstream.}
Train-synthetic, test-real segmentation with a compact U-Net
(skeleton-\gls{skiou} over \tstrNFolds\ recording-grouped folds): \synmtv (Ours)
matches the real-data upper bound and helps as augmentation, while the parametric
\synmtone\ transfers less of the real appearance.
Boxes: interquartile range, line: median, computed over \tstrNFolds\
recording-grouped folds.
Dashed line and grey band: the inter-annotator agreement on real images
(\gls{skiou}~\skiouHuman~$\pm$~\skiouHumanStd).}
\label{fig:tstr}
\end{figure}

\paragraph{Protocol.}
We train an \gls{mt} U-Net segmentation model on five data sources: (i) real images, (ii) the parametric baseline \synmtone, (iii) \synmtv, (iv) real$+$\synmtv, and (v) \emph{Naive}, samples from the raw last-epoch checkpoint of the same diffusion run without diagnostic-based selection.
The Naive arm isolates the effect of checkpoint selection from the effect of synthetic data quality.
All models are evaluated on held-out real images using recording-level splits to prevent leakage and share the same architecture, augmentation, and training budget.

\paragraph{Results.}
\Cref{fig:tstr} shows that synthetic data quality directly affects downstream utility.
Training on \synmtv reaches \gls{skiou}~\tstrVtwo\ on real images, approaching the real-data model (\tstrReal) and outperforming the parametric baseline (\tstrVone); combining real and \synmtv further improves performance to \tstrRealVtwo, ahead of real-only training in two of the four recording-grouped folds and behind it in the other two, a modest but consistent split-wise pattern given the small number of folds.
The \synmtv-trained model is also more stable across recording-grouped folds than the real-only model.

The Naive comparison directly tests the diagnostic: although generated from the same training run, the unselected final checkpoint reaches only \gls{skiou}~\tstrVlast, substantially below the diagnostic-selected model.
This demonstrates that checkpoint selection, rather than training duration alone, determines downstream utility.

\subsection{Repeated synthetic-to-real hyperparameter transfer}
\label{sec:hpo-transfer}

\paragraph{Protocol.}
Foundation segmenters can be adapted to a new modality by tuning a few inference hyperparameters.
We reproduce the hyperparameter-transfer experiment of \citet{koddenbrock2026synthetic}, who tuned the three segmenters strongest on synthetic-to-real transfer in that study (\hpoModels{}; of these, only SAM3Text reached ``human-grade'' analysis there) on a single $10$-image subset of \synmtone\ and transferred the result to real, reporting \gls{skiou}~$0.93$ on synthetic data but $0.74\pm0.18$ on real images.
A single $10$-image draw is a fragile basis for that estimate, so we repeat the tuning over \hpoNSeeds\ independent random subsets (\hpoNTrials\ optimization trials each) and report the distribution over subsets.
We run the identical protocol with both \synmtone\ and \synmtv\ as the tuning source, and evaluate on the same held-out real images, computing the same per-image mean \gls{skiou} they report for each of our \hpoNSeeds\ subsets before summarizing the resulting distribution by its median.

\paragraph{Results.}
The reproduction yields two findings.
First, results are better with our data: the median transfer per segmenter is \hpoVtwoMedLo--\hpoVtwoMedHi\ \gls{skiou} for \synmtv\ versus \hpoVoneMedLo--\hpoVoneMedHi\ for \synmtone\ (\synmtv\ is level or slightly ahead for all three segmenters; for example SAM3Text \hpoSamVtwo\ vs \hpoSamVone).
Second, the original single-subset result was a bit lucky: across \hpoNSeeds\ random subsets on \synmtone, SAM3Text's real \gls{skiou} ranges from \hpoSamVoneLo\ to \hpoSamVoneHi\ (median \hpoSamVone), and the study's reported $0.74$ sits near the top of that range.
These results show that \synmtv provides a more reliable tuning source, reducing the sensitivity of synthetic-to-real transfer to the choice of a small labeled subset.
SAM3Text reaches \gls{skiou}~\hpoSamVtwo\ when tuned on only $10$ images, comparable to the \tstrReal\ that a compact U-Net reaches only with $\sim\!150$ real tiles, so the remaining gap to full training is a matter of budget rather than method.

\paragraph{Comparison to real-data tuning.}
\citet{koddenbrock2026synthetic} also compared tuning SAM3Text on $10$ real or $10$ \synmtone\ images, reporting \gls{skiou}~$0.73$ and $0.71$ respectively, in a single-seed experiment.
In our evaluation across \hpoNSeeds\ tuning seeds, SAM3Text achieves median \gls{skiou}~\hpoSamVtwo\ with \synmtv\ and \hpoSamVone\ with \synmtone.
The published real-data result is therefore numerically close to our \synmtv\ median, although a single-seed result does not establish equivalence with our repeated evaluation, and the two protocols are not directly comparable.

\begin{figure}[tb]
\centering
\includegraphics[width=\linewidth]{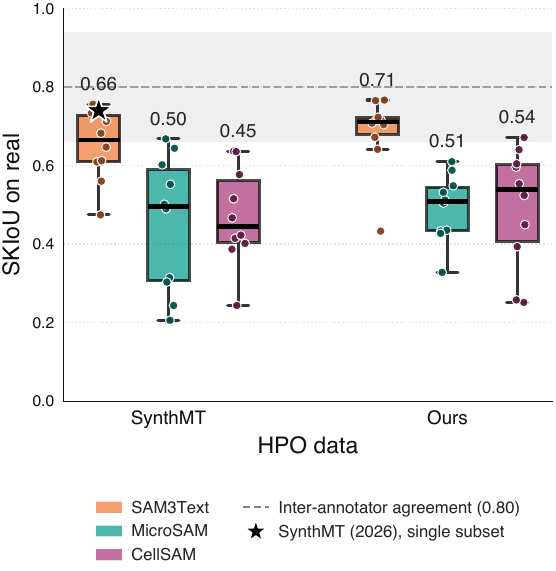}
\caption{\textbf{Repeated synthetic-to-real transfer of tuned segmenter
hyperparameters.} Each box summarizes \hpoNSeeds\ independent tuning seeds
(random $10$-image subsets, \hpoNTrials\ trials each) for one source and
segmenter (\hpoModels{}, the three models selected as the best
synthetic-to-real transfer segmenters in the SynthMT study,
\citealp{koddenbrock2026synthetic}); the spread exposes single-subset fragility, and the
\synmtv-vs-\synmtone offset measures the benefit of a more realistic tuning
source.
The black star marks the single-subset SAM3Text result originally reported
in \citet{koddenbrock2026synthetic} (\gls{skiou}~$0.74$): it sits near the top
of our repeated-seed distribution for the same (source, model) combination,
not at its median.
Boxes: interquartile range, line: median, computed over \hpoNSeeds\ tuning seeds.
Dashed line and grey band: the inter-annotator agreement on real images
(\gls{skiou}~\skiouHuman~$\pm$~\skiouHumanStd).}
\label{fig:hpo-transfer}
\end{figure}

\Cref{fig:hpo-transfer} shows the per-seed transfer distributions for each source and segmenter.

\section{Discussion}
\label{sec:discussion}

The failure mode identified in this work has a simple mechanistic explanation.
When conditioning carries most of the spatial information, as in mask-conditioned diffusion, the remaining appearance degrees of freedom (texture, background, brightness) can collapse onto low-risk attractors while preserving the mask-defined structure.
Late in training, we observe this as convergence toward a narrow color palette: the anti-correlation between color distance and inter-similarity ($r=-0.71$, \cref{fig:binned-corr}) indicates that increasing similarity is achieved at the expense of appearance diversity.
A related mechanism explains early collapse, where samples remain close to the dataset mean while still occupying the reference distribution in DINOv2 space.
We expect this failure mode to extend beyond microtubule microscopy to other spatially conditioned generators, including pixel-space diffusion and small-domain \glspl{cgan} such as SPADE~\citep{park2019spade}.

The aggregation strategy in \cref{eq:agg} reflects this observation.
A geometric mean prevents a strong score on one axis from compensating for collapse on another.

The baseline-aware interpretation of mask fidelity highlights a broader issue in evaluating synthetic biomedical images.
For thin filaments, segmentation itself is ambiguous: experts agree only to \gls{skiou}~\skiouHuman\ on real images.
Therefore, geometric fidelity must be interpreted relative to this achievable ceiling.
The comparison in \cref{app:comparison} shows that a synthetic system can exceed this ceiling without necessarily producing more realistic images.

\paragraph{Limitations.}
Our evaluation is based on a relatively small dataset ($309$ labeled $512\times512$ crops) from a single biomedical domain.
Although \synmtv shows downstream utility, the train-on-synthetic experiment uses real expert masks as conditioning, whereas the parametric baseline relies on procedural masks; thus, differences in mask distribution may contribute alongside appearance realism.
Separating these two effects would require a controlled comparison on matched conditioning masks, which we leave to future work; likewise, generating synthetic masks that capture realistic microtubule geometry, so that the pipeline no longer depends on real expert masks at inference time, is nontrivial and is itself a direction for future work.
The hyperparameter-transfer experiment demonstrates improved adaptation over the parametric synthetic baseline, but does not establish superiority over tuning directly on an equally sized real subset, since \synmtv itself is generated from a model trained on real microscopy.
Finally, while the mechanism likely extends to other spatially conditioned generators, demonstrating the same failure mode across additional biomedical domains such as \gls{ct} and \gls{mri} remains future work.
\section{Conclusion}
\label{sec:conclusion}

We introduce \synmtv, an openly available triplet dataset of pixel-precise masks, real \gls{irm} microscopy images, and mask-conditioned synthetic counterparts.
Unlike standard generative metrics, our diversity-aware diagnostic identifies the collapse of unconstrained appearance features that can occur while realism scores remain high.
The selected model produces synthetic images that domain experts cannot reliably distinguish from real recordings in a 2AFC study and generates training data that transfers to real segmentation: a model trained only on \synmtv reaches segmentation performance comparable to a real-trained model and substantially exceeds the parametric baseline.
Beyond this dataset, our diagnostic provides a general evaluation tool for mask-conditioned diffusion models, where realism and diversity must be assessed jointly rather than through a single distributional score.

\iffinal
\paragraph{Data and code.}
\synmtv, the trained checkpoint, the 2AFC challenge, and the diagnostic implementation are publicly available.
The diagnostic is released as a standalone tool that can be applied to other mask-conditioned diffusion models.
All resources are available at:
\url{https://huggingface.co/spaces/HTW-KI-Werkstatt/DiffuMT}.
\else
\paragraph{Data and code.}
The dataset, trained checkpoint, 2AFC challenge, and interactive browser demo will be released publicly upon acceptance.
The diagnostic implementation will be released as a standalone tool applicable to other mask-conditioned diffusion models.
\textit{Repository URLs are withheld here to preserve double-blind review.}
\fi

\iffinal
\section*{Acknowledgments}

Our work is funded by the Deutsche Forschungsgemeinschaft (DFG, German Research Foundation) Project-ID 528483508 -- FIP 12.
We would like to thank the Reber lab at the Advanced Medical Bioimaging Core Facility (Charit\'e, Berlin) for imaging support, and its members who served as expert participants in the 2AFC study.

\fi
\section*{LLM Disclosure}

In this work, we used general-purpose LLM/AI assistants for implementation, \ie writing and debugging code.
We have not used LLMs/AI tools for synthetic data, theoretical or conceptual frameworks, mathematical claims or proofs, hypotheses, methodology or experimental design, translation, data cleaning, qualitative data analysis, or interpreting results; these tasks are not applicable to AI assistance in this work.
Additionally, we used LLMs/AI tools for drafting and editing parts of the paper to improve grammar and phrasing.
We have reviewed all AI-assisted work: all authors read and verified the generated code against our own experiments and checked every AI-edited sentence against the original draft and our intended meaning.
We take responsibility for the final content of this work, including text, claims, or artifacts produced with the aid of LLMs/AI tools.

\printbibliography

\appendix
\section{Parametric vs.\ generative baseline comparison}
\label{app:comparison}


The diagnostic of \cref{sec:method} is most useful when it changes the conclusion of a comparison that a single metric would settle incorrectly.
We illustrate this with a head-to-head comparison of the segmentation-guided \gls{dm} against the parametric baseline \synmtone~\citep{koddenbrock2026synthetic} on identical synthetic masks.

\paragraph{Setup.}
We sample $1000$ images from each model, conditioned on the same synthetic mask set (the one released with~\citet{koddenbrock2026synthetic}), and evaluate all with the diagnostic of \cref{sec:diag}.
We additionally sample the \emph{Naive} arm (\cref{sec:tstr}: the raw last-epoch checkpoint, bypassing the checkpoint-selection procedure) on the same masks, so all four systems are directly comparable.
The reference for the similarity and color metrics is the held-out real test set used to tune the \gls{dm}; the reference for the geometric metrics is the conditioning mask itself.

\paragraph{Results.}
\Cref{tab:comparison} (\cref{sec:experiments}) reports the head-to-head comparison.
On the geometric mask-fidelity rows, \synmtone scores higher in absolute terms (\gls{skiou}~$0.96$ vs.\ $0.86$, F1@75~$0.97$ vs.\ $0.88$), but this is by construction, since it paints filaments directly along the polylines used to build the mask.
The two \gls{skiou} readings for \synmtv, $0.86$ here and $0.67$ in \cref{tab:absolute-runs}, come from the same model on different mask sets: $0.86$ on the $1000$ synthetic masks of this comparison and $0.67$ on the $200$ held-out real masks.
Both sit at the inter-annotator band, whereas \synmtone's $0.96$ exceeds it on either set.
Read relative to the inter-annotator agreement on real images (\gls{skiou}~\skiouHuman), \synmtv is the closer match on either mask set (a gap of $0.06$ on the synthetic masks here, $0.13$ on the real masks of \cref{tab:absolute-runs}), whereas the parametric model sits $0.16$ above the human ceiling and is thus easier to segment than reality.
On every axis where the comparison is informative, \synmtv is the more realistic system: inter-similarity~$0.97$ vs.\ $0.86$, diversity ($\Delta$Intra-Sim.\ $0.01$ vs.\ $0.06$), distribution match (\gls{dfid}~$0.14$ vs.\ $0.27$), and color (global \gls{emd}~$3.56$ vs.\ $8.93$).
The Naive arm, evaluated on the same masks, makes the diagnostic's point directly: its mask fidelity is indistinguishable from \synmtv's (\gls{skiou}~$0.86$ for both) and it scores the best \gls{dfid} of the three DM/parametric systems ($0.07$), yet its diversity gap is five times wider than \synmtv's ($\Delta$Intra-Sim.\ $0.05$ vs.\ $0.01$) and its color fidelity is worse overall and in the foreground, though marginally better in the background.
\gls{dfid} improves exactly where the diagnostic predicts it would --- at the expense of diversity.

\paragraph{Implication for downstream use.}
Because \synmtone's \gls{skiou} lies above the inter-annotator ceiling, it has not reproduced the segmentation difficulty that downstream models encounter on real microscopy.
Whether this is desirable depends on the task, but the diagnostic surfaces it, whereas an absolute reading would not.

\begin{figure*}[tb]
\centering
\includegraphics[width=\textwidth]{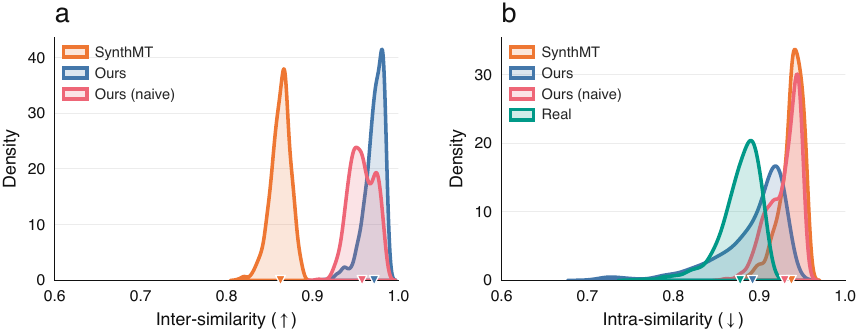}
\caption{\textbf{Our \gls{dm} matches the real distribution on both axes; \synmtone is shifted left on realism (left) and collapsed onto a narrow high-similarity band on diversity (right).}
Density estimates of inter- and intra-similarity against the real reference.
The naive arm (raw last-epoch checkpoint, no diagnostic selection, \cref{sec:downstream}) is shown for reference: it matches \synmtv on realism (inter-similarity) but collapses toward \synmtone's narrow, high-similarity footprint on diversity (intra-similarity) -- individually realistic yet selectively losing diversity.}
\label{fig:kde}
\end{figure*}

\begin{figure*}[tb]
\centering
\includegraphics[width=\linewidth]{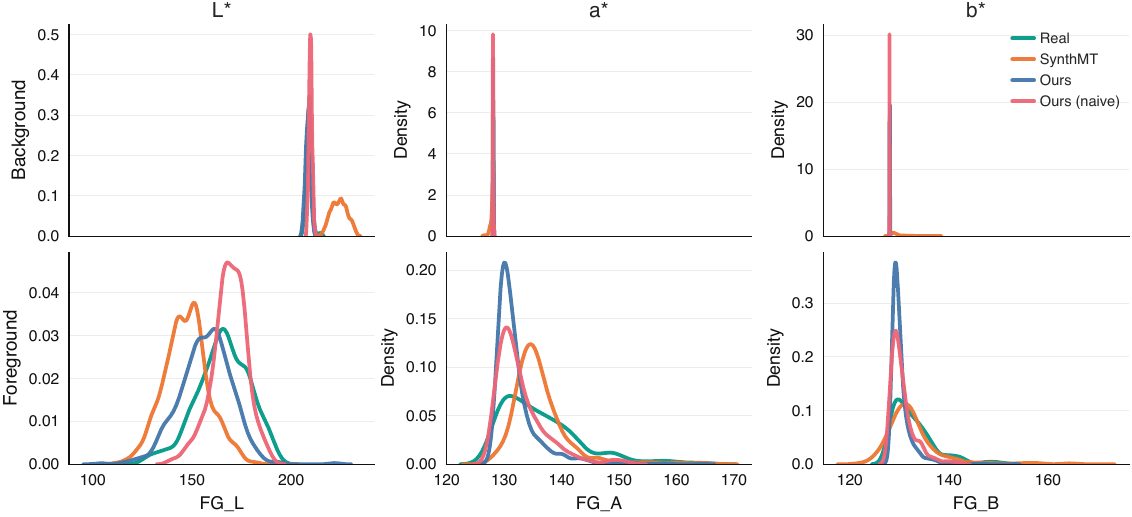}
\caption{\textbf{Our \gls{dm} reproduces the real \gls{irm} background distribution; the parametric \synmtone baseline does not.}
\gls{cielab} background (top) and foreground (bottom) distributions for \synmtone, our \gls{dm} and the real reference.
The \gls{dm} background overlaps the reference; \synmtone's is visibly too bright.
In the foreground both fall within a plausible range, with the \gls{dm} slightly over-concentrated.}
\label{fig:cielab-cmp}
\end{figure*}

\begin{figure*}[p]
\centering
\includegraphics[width=\linewidth]{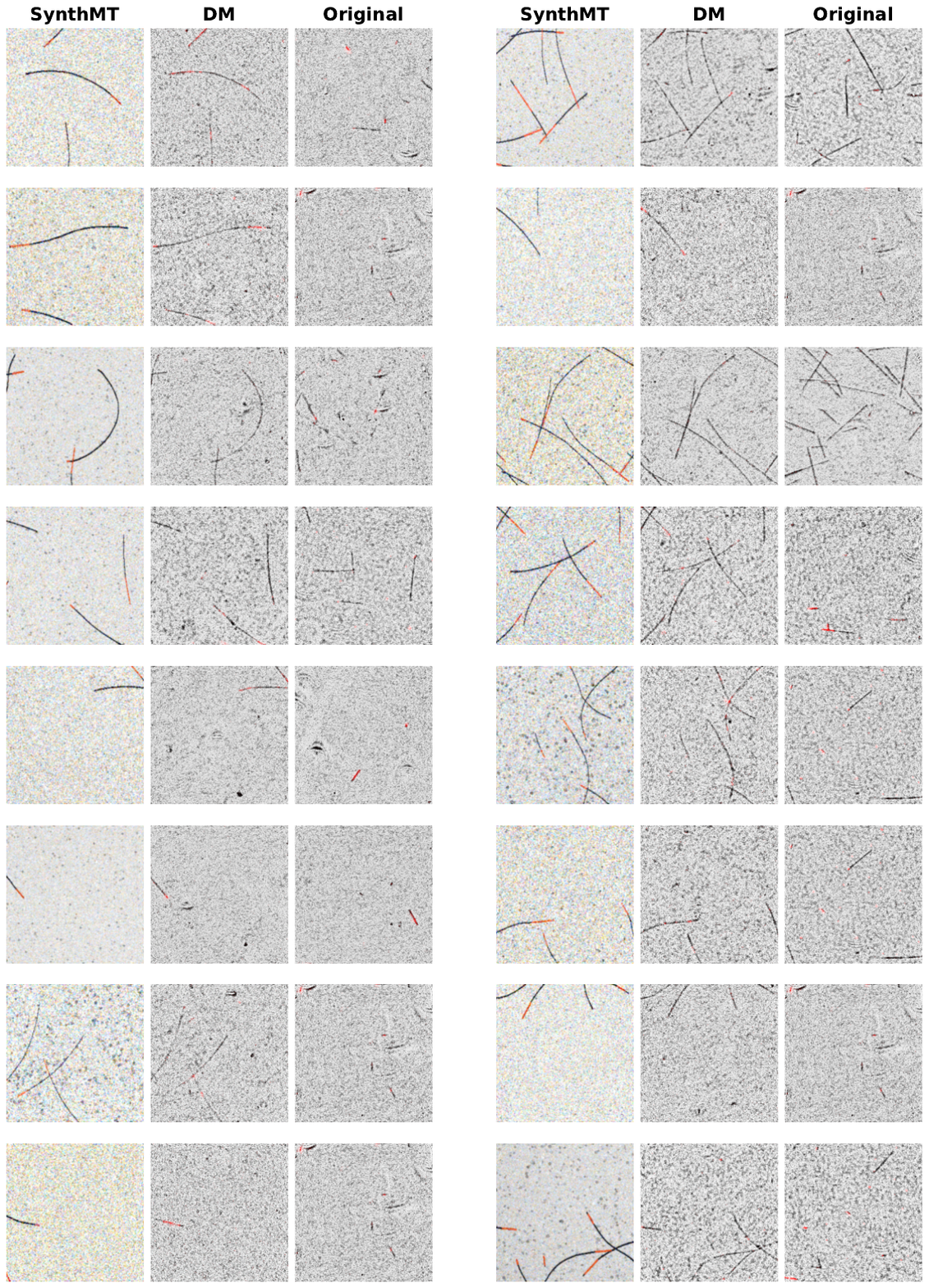}
\caption{\textbf{Our \gls{dm} recovers camera noise, \gls{irm} background gradients and irregular filament brightness that the parametric baseline cannot reproduce, at the cost of looser geometric mask adherence.}
Sixteen triplets, each row left to right: a \synmtone parametric sample, our \synmtv sample on the same synthetic mask, and the real image most similar to it (nearest neighbor, $k\!=\!1$).}
\label{fig:triplets}
\end{figure*}

\paragraph{Distributional and qualitative evidence.}
\Cref{fig:kde} shows the same conclusions distributionally.
The \gls{dm}'s intra-similarity overlaps the real distribution, matching the natural diversity, while \synmtone's sits high and narrow, the expected footprint of a parametric pipeline that re-samples from a small parameter family.
The naive arm, plotted alongside for reference, stays on the realistic side on inter-similarity but drifts back toward \synmtone's collapsed footprint on intra-similarity, consistent with the downstream utility drop quantified in \cref{sec:downstream}: realism alone does not guarantee diversity survives past the diagnostic-selected checkpoint.
The \gls{cielab} comparison (\cref{app:additional}, \cref{fig:cielab-cmp}) adds a color-based view of the same three systems.
Sixteen qualitative triplets (synthetic mask, \synmtone, \gls{dm}, nearest real neighbor) are shown in \cref{app:additional}, \cref{fig:triplets}.

\section{Ablation comparison across configurations}
\label{app:ablation}

\Cref{tab:absolute-runs} lists absolute metric values for a few representative runs, each at its own diagnostic-selected checkpoint (\cref{sec:diag}, applied uniformly across all runs rather than fixing an epoch), all evaluated on the same $200$ held-out real masks (for the mask-fidelity and similarity axes).
The \textit{Low-data} run, trained on only 50 labeled images, attains the highest inter-similarity (\textbf{0.98}) of any run in the paper, yet its intra-similarity (\textbf{0.97}) reveals a sharp mode collapse that inter-similarity hides.
The \textit{Aug} run (MSE replaced by Huber, with Gaussian input noise) shows the opposite pattern: it scores best on mask fidelity but loses foreground color fidelity by a factor of about $3$ relative to the \synmtv configuration.
The \synmtv run is the only one of the three that does not have a severe outlier on any single axis in the table, and its samples form the released dataset.

\begin{table}[tb]
\centering
\caption{\textbf{\synmtv does not collapse on any axis, whereas the contrastive runs each collapse on diversity (Low-data) or overfit the segmenter (Aug).}
\textbf{Real}: inter-annotator agreement (\skiouHuman) for mask fidelity, real-test statistics otherwise; not a candidate model.
``--'' = no real-vs-real value; \textbf{bold} = best of the three \gls{dm} columns per row.}
\label{tab:absolute-runs}
\resizebox{\linewidth}{!}{%
\begin{tabular}{lcccc}
\toprule
& \textbf{Real} & \multicolumn{3}{c}{\textbf{Our mask-conditioned \gls{dm}}} \\
\cmidrule(lr){2-2}\cmidrule(lr){3-5}
\textbf{Metric} & \textbf{(baseline)} & \synmtv & \textbf{Low-data} & \textbf{Aug} \\ \midrule
\multicolumn{5}{l}{\textit{Mask fidelity}} \\
\gls{skiou} ($\uparrow$)                & $0.80$ & $0.67$ & $0.62$ & $\mathbf{0.70}$ \\
F1@50 ($\uparrow$)                & $0.89$ & $\mathbf{0.74}$ & $0.70$ & $\mathbf{0.74}$ \\
F1@75 ($\uparrow$)                & $0.87$ & $0.66$ & $0.60$ & $\mathbf{0.70}$ \\
\midrule
\multicolumn{5}{l}{\textit{Distribution match}} \\
\acrshort{dfid} ($\downarrow$)                & --    & $0.11$ & $0.17$ & $\mathbf{0.07}$ \\
\midrule
\multicolumn{5}{l}{\textit{Color}} \\
Color, foreground ($\downarrow$) & --    & $\mathbf{1.18}$ & $7.01$ & $4.10$ \\
Color, background ($\downarrow$) & --    & $1.88$ & $\mathbf{0.76}$ & $2.32$ \\
\midrule
\multicolumn{5}{l}{\textit{Similarity}} \\
Inter-Sim. ($\uparrow$)           & --    & $0.97$ & $\mathbf{0.98}$ & $0.97$ \\
\midrule
\multicolumn{5}{l}{\textit{Diversity}} \\
Intra-Sim. ($\downarrow$)         & $0.88$ & $\mathbf{0.88}$ & $0.97$ & $0.89$ \\
\bottomrule
\end{tabular}}
\end{table}

\section{Implementation details}
\label{app:impl}

\paragraph{Architecture.}
2D U-Net from the segmentation-guided diffusion implementation of \citet{konz2024}: four down-/up-sampling stages, residual blocks, channel multipliers $(1, 2, 4, 8)$ on a base of 128 channels, sinusoidal time embedding, one extra input channel for the binary mask.
About $113$\,M parameters total.

\paragraph{Training.}
AdamW with $\beta_1\!=\!0.9$, $\beta_2\!=\!0.999$, weight decay $10^{-2}$; learning rate $10^{-4}$ with a cosine schedule and $500$ warm-up steps; mixed precision (fp16); \gls{ema} on the U-Net weights with decay $0.9999$; batch size $16$; $1000$ forward diffusion steps with a linear $\beta$-schedule (one ablation uses scaled cosine).
Each run uses $186$ steps per epoch; the hyperparameter ablations run for $400$ or $600$ epochs, and the full diagnostics sweep of \cref{app:agg-robustness,app:checkpoints} extends to $1000$ epochs to characterize late-training behavior.
A full $600$-epoch run takes about 18 hours on a single NVIDIA A100.

\paragraph{Sampling.}
\gls{ddim}~\citep{song2021ddim} with $50$ inference steps and $\eta\!=\!0$ (deterministic).
The binary conditioning mask is fed in the same dynamic range as during training---label values $\{0,1\}$ divided by $255$ by the dataloader, i.e.\ $\{0,\,1/255\}$---since feeding $\{0,1\}$ directly is a $255\times$ stronger conditioning signal that pushes the model out of distribution.
Generating $1000$ patches on a consumer NVIDIA RTX~5070~Ti ($16$\,GB VRAM) takes about $17$ minutes.

\paragraph{Evaluation hyperparameters.}
DINOv2 ViT-B/14, fifth block, $L_2$-normalized CLS token.
$k\!=\!10$ nearest neighbors for inter-similarity.
$200$ samples per evaluated checkpoint; $1000$ samples for the parametric comparison.
\gls{dkid} polynomial kernel of degree $3$, matching the \synmtone release.
Foreground/background re-segmentation with SAM~3~\citep{sam32025} in its text-prompted mode (prompt \texttt{"thin line"}), following \citet{koddenbrock2026synthetic}.

\paragraph{Aggregation.}
The global selection score $S_\text{global}$ (\cref{eq:agg}) is the geometric mean of three min-max-scaled perceptual axes: inter-similarity, intra-similarity and color distance.
For \emph{checkpoint selection within a run}, each axis is scaled within that run so the best checkpoint scores $1.0$ and the worst $0.0$; the gap-style axis (intra-similarity) is scored against the real-test value before scaling.
\Cref{tab:absolute-runs} reports raw per-axis values rather than a cross-run $S_\text{global}$ column, since min-max scaling three runs against each other is only meaningful at the full precision of the underlying measurements, not the values as displayed; the within-run selection of epoch~\runFifteenEpoch{} does not depend on this table.
Geometric mask-fidelity axes and \gls{dfid} are reported per-checkpoint but excluded from $S_\text{global}$ (see \cref{eq:agg} and the rationale in \cref{sec:diag}).

\section{Metric correlations: why no single metric suffices}
\label{app:binned-corr}

\begin{figure*}[tb]
\centering
\includegraphics[width=\textwidth]{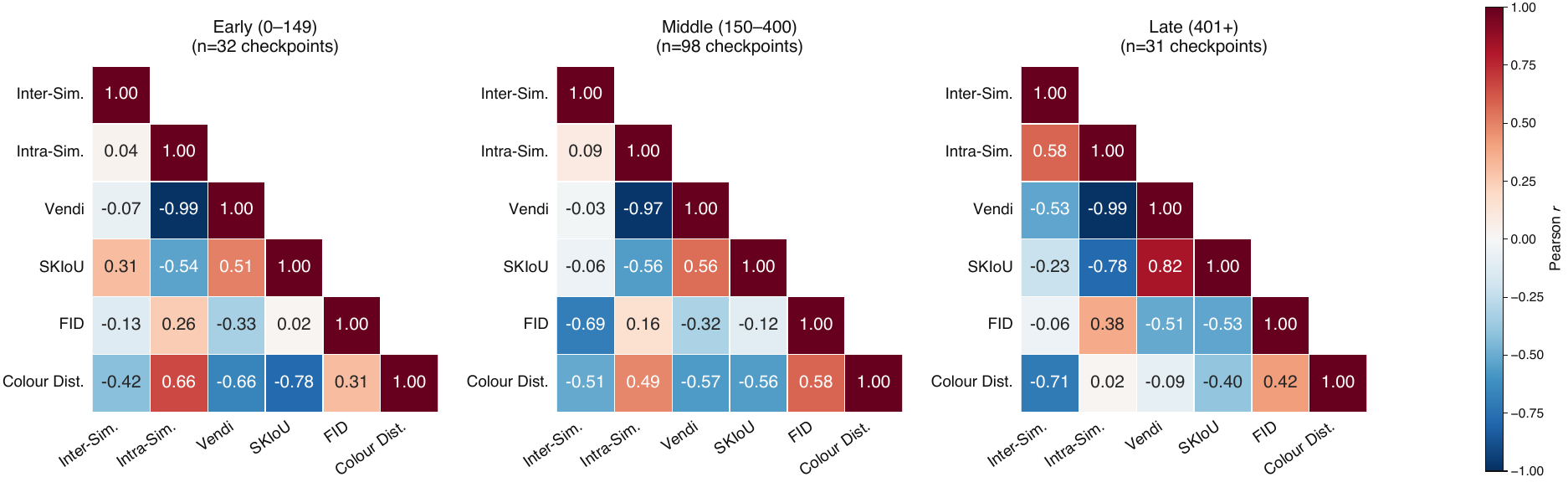}
\caption{\textbf{Metric correlations reveal the failure mode: in late training, inter- and intra-similarity re-couple positively while color distance becomes strongly anti-correlated with inter-similarity, indicating a texture-diversity collapse the standard toolbox misses.}
Pearson correlations between evaluation metrics within three epoch bins (early: 0--149 / middle: 150--400 / late: $>$400), pooled over all training runs. The late-regime inter-vs-intra re-coupling (r = +0.58) is the core observation justifying the diagnostic; the color--similarity anti-correlation (r = --0.71) corroborates it. The Vendi score is included alongside intra-similarity and tracks it almost perfectly in every bin, consistent with \cref{app:vendi}.}
\label{fig:binned-corr}
\end{figure*}

\Cref{fig:binned-corr} resolves the numerical signature of \cref{sec:exp-degeneracy} into the full pairwise correlation structure across the three epoch bins, which is the quantitative reason that no single metric suffices.
The axes are not interchangeable.
In the early and middle bins, realism and diversity are decoupled (inter- vs.\ intra-similarity $r\!=\!+0.09$ mid-training), so neither can serve as a proxy for the other.
In the late bin they conflict.
The same pair re-couples positively ($r\!=\!+0.58$): realism keeps rising while diversity falls, which is the collapse itself, not merely a correlate of it.
Color distance corroborates the same picture, becoming strongly anti-correlated with inter-similarity ($r\!=\!-0.71$; $n\!=\!31$ late-bin checkpoints, $95\%$ Fisher-$z$ CI $[-0.85,-0.48]$, $p<0.001$) as the model settles onto a fixed palette; \gls{dfid} tracks inter-similarity rather than diversity, so it improves through the same window.
The Vendi score~\citep{friedman2023vendi}, included here for the same reason as \cref{app:vendi}, tracks intra-similarity almost perfectly in every bin ($|r|\!\geq\!0.97$), so it moves with the collapse rather than flagging it independently.
Each axis therefore has a degenerate optimum of its own: a realism- or \gls{dfid}-only score is maximized by a low-diversity mode collapse, while a diversity-only score is maximized by incoherent samples that ignore the data.
Because color moves \emph{with} inter-similarity rather than against it in this regime, it does not itself flag the late-training collapse; but $S_\text{global}$ is a geometric mean (\cref{eq:agg}), so a metric that fails to flag a collapse cannot rescue a score that the diversity axis has already driven toward zero.
The diagnostic instead reads a realism/distribution anchor together with a diversity axis, where the anchor keeps the diversity axis from rewarding noise and the diversity axis keeps the anchor from rewarding collapse.
The same late-bin collapse also makes the final training epoch the wrong checkpoint (\cref{app:checkpoints}), which is a corollary of this structure.

\section{Intra-similarity agrees with the Vendi score}
\label{app:vendi}

The diversity axis of our diagnostic is the mean pairwise cosine intra-similarity (\cref{sec:diag}).
We use it rather than an off-the-shelf diversity score because it lies on the same DINOv2 cosine scale as the inter-similarity (realism) axis, so realism and diversity are read on one comparable scale and against the same real-data band.
To confirm that this sacrifices no rigor, we computed the Vendi score~\citep{friedman2023vendi}, the standard reference-free diversity metric, on the same embeddings for every checkpoint of the eight hyperparameter-ablation runs ($n\!=\!161$ checkpoints).
Two results follow.
First, intra-similarity and the Vendi score are almost perfectly (inversely) correlated (Pearson $|r|\!=\!0.98$, Spearman $\rho\!=\!-0.98$), and both place the diagnostic-selected checkpoint at peak diversity and the late-training checkpoints in collapse.
Second, and more directly, replacing the intra-similarity axis with a Vendi-gap axis in $S_\text{global}$ selects the identical checkpoint on every one of the eight hyperparameter runs; on \synmtv the swap also leaves the selected checkpoint at epoch~\runFifteenEpoch{} unchanged.
The diagnostic is therefore robust to this choice.
We keep intra-similarity because it is cheaper and lies on the same scale as the realism axis, and report the Vendi score only as this cross-check, since including both would merely double-count diversity.

\section{Sensitivity to the offset-noise scale}
\label{app:sigma-robustness}

\begin{figure*}[tb]
\centering
\includegraphics[width=\linewidth]{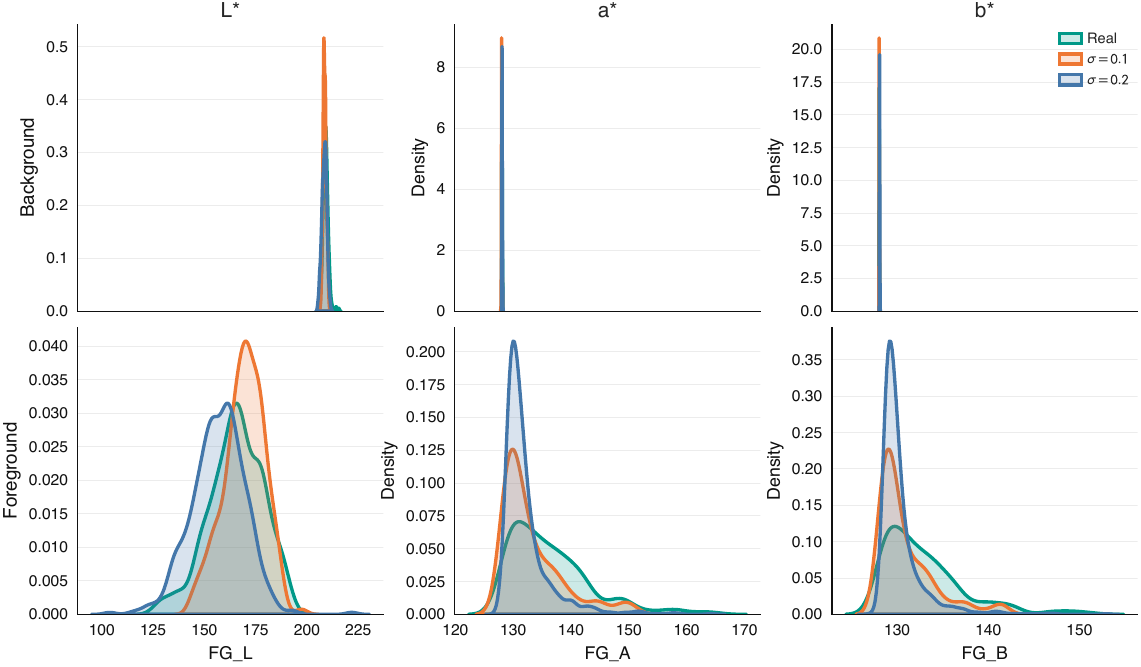}
\caption{\textbf{Both offset-noise settings ($\sigma\!=\!0.1, 0.2$) reproduce the real \gls{cielab} distribution; $\sigma\!=\!0.2$ matches the real background variance slightly more faithfully.}
\gls{cielab} background (top) and foreground (bottom) distributions for the real reference and our \gls{dm} trained at the two offset scales.}
\label{fig:cielab-sigma}
\end{figure*}

The main \synmtv runs use offset noise with $\sigma\!=\!0.2$.
A companion run at $\sigma\!=\!0.1$ (otherwise identical recipe) produces a visually similar \gls{cielab} distribution (\cref{fig:cielab-sigma}): both variants concentrate the background luminance near the real-data mode, and both spread the foreground luminance across the same range.
$\sigma\!=\!0.2$ reproduces the real background variance slightly more faithfully ($\text{std}\,\text{BG\_L}\!=\!1.21$ vs.\ $0.77$ for $\sigma\!=\!0.1$; real reference $1.35$), which is the basis for using it throughout the main paper.
The diagnostic-selected epoch does not change between the two settings.

\section{Robustness of the selected checkpoint}
\label{app:agg-robustness}

A natural question is whether the choice of the geometric mean is critical.
\Cref{tab:agg-robustness-components} reports the checkpoint selected (argmax) by the geometric mean used in the main paper (\cref{eq:agg}), the arithmetic mean of the same three axes, a diversity-only score (the intra-similarity axis $\hat s_\text{Intra}$ alone), and every single-axis and pairwise-axis reduction of $S_\text{global}$, all over the full $200$-checkpoint training trajectory of \synmtv (1000 epochs, every-5th-epoch diagnostics sweep).
The geometric and arithmetic means agree, both selecting epoch~\runFifteenEpoch{}; the diversity-only score selects a nearby but distinct epoch~\runFifteenDiversityEpoch{}.
This is weaker than full three-way agreement, but the two aggregations that combine both selection axes converge on the same checkpoint, which is the property the diagnostic depends on: the aggregation rule matters less than reading diversity jointly with realism.
\gls{dfid} is still reported but excluded from $S_\text{global}$ for the reasons of \cref{sec:method}: it conflates quality and diversity into one number, so it cannot be relied upon to flag the collapse of \cref{sec:exp-degeneracy} as reliably as the joint realism/diversity score does (\cref{app:checkpoints}).

\paragraph{Which axis does the work?}
Either axis alone is unreliable: realism-only selects epoch~$90$, well before the collapse of \cref{sec:exp-degeneracy} even begins, and color-only selects epoch~$350$, a different point again.
Dropping color and keeping realism and diversity jointly (geometric mean of the two) still selects epoch~\runFifteenEpoch{}, the same checkpoint as the full three-axis score, so color is not what drives the selection.
Dropping realism and keeping diversity and color instead selects epoch~\runFifteenDiversityEpoch{}, the same as the diversity-only score, so with realism removed the choice is governed entirely by diversity.
Realism and diversity read jointly are therefore what the selection depends on; color is corroborating rather than decisive.

\begin{table}[tb]
\centering
\caption{\textbf{Single- and pairwise-axis ablations of the selection score all select an epoch within the same narrow window as the full three-axis score, except for the two single-axis scores.} Selected epoch (argmax) for each reduction of $S_\text{global}$ on the run15\_B16 trajectory.}
\begin{tabular}{lc}
\toprule
\textbf{Aggregation scheme} & \textbf{Selected epoch} \\
\midrule
Geometric mean (paper) & 315 \\
Arithmetic mean & 315 \\
Diversity-only (intra-sim) & 250 \\
Realism-only (inter-sim) & 90 \\
Color-only & 350 \\
Realism+diversity (no color) & 315 \\
Diversity+color (no realism) & 250 \\
\bottomrule
\end{tabular}

\label{tab:agg-robustness-components}
\end{table}

\section{Corrected reference pool for checkpoint selection}
\label{app:corrected-reference}

The main text (\cref{sec:twoafc}) reports that our original checkpoint-selection reference pool for the Inter-similarity and Colour axes of \cref{eq:agg} was not fully held out from training: approximately $58\%$ of the similarity-pool images, and nearly all of the Colour-pool images, were also training images. We thank the reviewer who identified this. We rebuilt both pools from a disjoint, held-out set of $200$ images and recomputed the selection score for all $200$ saved checkpoints; under this correction, epoch~\runFifteenEpoch{} ranks second (score $0.7433$) instead of first, behind epoch $350$ (score $0.7855$), a gap small enough to plausibly reflect sampling noise from the rerun's freshly generated samples. The two robustness checks below use this corrected pool.

\section{Robustness to the choice of embedding model}
\label{app:model-ablation}

Holding the aggregation rule fixed, we repeated the corrected-pool selection (\cref{app:corrected-reference}) with four other embedding models substituted for DINOv2 in the Inter and Intra axes: ResNet-50, CLIP, MorphEm, and Cell-DINO. \Cref{tab:model-ablation} reports the top-ranked epoch under each. ResNet-50, MorphEm, and Cell-DINO agree with DINOv2 on the top-ranked epoch; CLIP selects a different epoch, consistent with its more object-focused representation~\citep{stein2023exposingflaws}. Agreement on the top epoch does not mean the complete rankings agree: under DINOv2, epoch~\runFifteenEpoch{} ranks $2$nd of $200$, but it ranks only $15$th under ResNet-50, $74$th under Cell-DINO, and $112$th under MorphEm, so this is not evidence of full representation-independent agreement.

\begin{table}[tb]
\centering
\caption{\textbf{Three of four alternative encoders select the same top-ranked epoch as DINOv2.} Corrected-reference selection (\cref{app:corrected-reference}) under each encoder, out of $200$ checkpoints.}
\label{tab:model-ablation}
\begin{tabular}{lr}
\toprule
Encoder & Top-ranked epoch \\
\midrule
DINOv2 (paper) & $350$ \\
ResNet-50 & $350$ \\
Cell-DINO & $350$ \\
MorphEm & $350$ \\
CLIP & $485$ \\
\bottomrule
\end{tabular}
\end{table}

\section{Robustness to the DINOv2 layer}
\label{app:layer-ablation}

We repeated the DINOv2-based selection (\cref{app:corrected-reference}) using features from layers $1$, $3$, $5$, $7$, $9$, and $11$ instead of the layer used elsewhere in the paper (\cref{sec:diag}), keeping the pooling method unchanged. Four of the six layers agree with the layer used throughout the paper; the earliest and deepest layers select different epochs. \Cref{tab:layer-ablation} also reports the selection without the Colour axis at each layer, which diverges from the full-diagnostic selection at most layers, confirming that the layer-5 agreement concerns the complete Inter/Intra/Colour score rather than the DINOv2 embedding alone.

\begin{table}[tb]
\centering
\caption{\textbf{Agreement with the layer-5 selection used throughout the paper holds at four of six tested layers, but only for the complete Inter/Intra/Colour score.} Corrected-reference selection (\cref{app:corrected-reference}) at each DINOv2 layer, with and without the Colour axis.}
\label{tab:layer-ablation}
\begin{tabular}{lrr}
\toprule
Layer & Full score & Without Colour \\
\midrule
$1$ & $830$ & $830$ \\
$3$ & $350$ & $430$ \\
$5$ (paper) & $350$ & $480$ \\
$7$ & $350$ & $295$ \\
$9$ & $350$ & $290$ \\
$11$ & $290$ & $275$ \\
\bottomrule
\end{tabular}
\end{table}

\section{The final epoch is not the right checkpoint}
\label{app:checkpoints}

\begin{figure*}[tb]
\centering
\includegraphics[width=\linewidth]{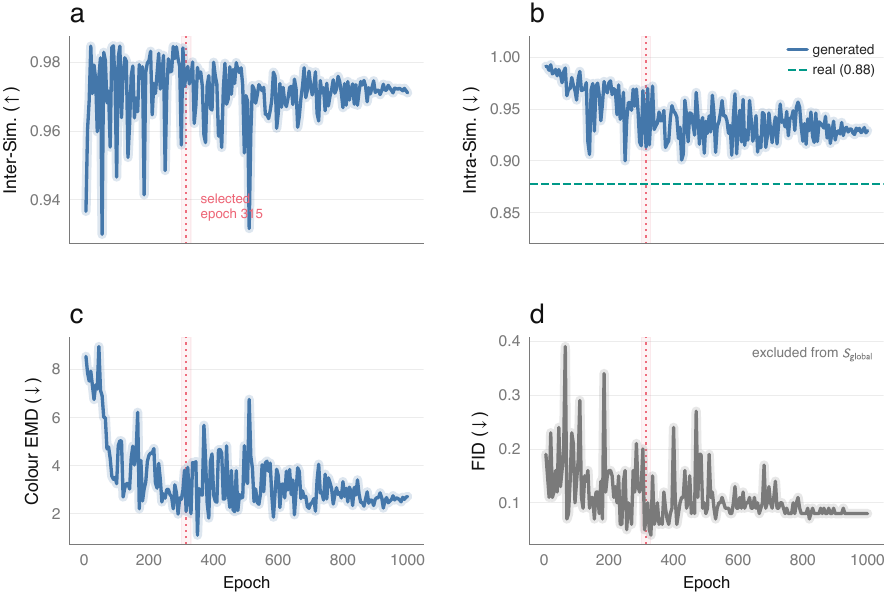}
\caption{\textbf{The diagnostic-selected checkpoint (epoch~\runFifteenEpoch{}) balances realism, diversity and color jointly; no single axis peaks there in isolation.}
Per-epoch trajectories of the three $S_\text{global}$ axes (inter-sim,
intra-sim, color) and \gls{dfid} for the \synmtv run; \gls{dfid} is
reported but excluded from the selection score. Dashed green: real-test
reference where defined; dotted red: selected checkpoint.}
\label{fig:epoch-trajectories}
\end{figure*}

\Cref{fig:epoch-trajectories} shows the three $S_\text{global}$ axes (inter-similarity, intra-similarity, color) and \gls{dfid} over training for the full \synmtv training run (1000 epochs, every-5th-epoch diagnostics sweep).
Inter-similarity plateaus at $\sim\!0.97$ from epoch~100 onward and is uninformative.
Intra-similarity, the diversity axis, is not at its own minimum at the selected checkpoint, epoch~\runFifteenEpoch{} (its minimum is nearby, epoch~\runFifteenDiversityEpoch{}); the geometric mean of \cref{eq:agg} weighs it jointly with inter-similarity and color, so the two selections differ slightly (\cref{app:agg-robustness}).
Because the geometric mean collapses when any axis falls, the diversity axis's post-peak degradation is what keeps $S_\text{global}$ from drifting to a later epoch even while inter-similarity keeps improving.
\gls{dfid} (panel d) stays flat and noisy through the same late-training window rather than rising with the diversity collapse; it conflates quality and diversity into one number rather than separating them (\cref{sec:related}), so we exclude it from $S_\text{global}$ on principle.
Color tracks the selection score without overriding the diversity signal, so the \{inter, intra, color\} score still selects epoch~\runFifteenEpoch{}.

\begin{figure}[tb]
\centering
\includegraphics[width=\linewidth]{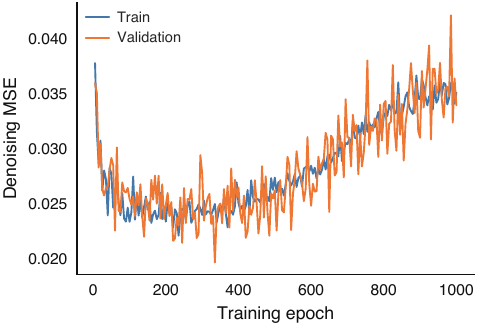}
\caption{\textbf{Training and validation denoising MSE closely track each other throughout training, with no widening generalization gap.} Both losses decrease rapidly during early training, reach their minimum after approximately one third of the training run, and subsequently increase together, indicating a deterioration of the denoising objective rather than conventional overfitting. Losses are Monte Carlo estimates based on eight timestep draws and four noise seeds per image, evaluated at each of the $200$ saved checkpoints.}
\label{fig:denoising-loss}
\end{figure}

\paragraph{Training vs.\ validation denoising loss.}
Because the diversity collapse of \cref{sec:exp-degeneracy} occurs late in training, we checked whether it coincides with classic overfitting. We computed training and validation denoising MSE at all $200$ saved checkpoints (eight timestep draws, four noise seeds per image); \cref{fig:denoising-loss} shows both curves. The two curves correlate at $0.87$ with no widening gap late in training, so this measurement does not show the signature of validation loss diverging from training loss; it does not, however, rule out other forms of memorization.

\section{2AFC study details}
\label{app:2afc}

The 2AFC validation reported in \cref{sec:twoafc} uses the HuggingFace Space (\iffinal\url{https://huggingface.co/spaces/HTW-KI-Werkstatt/DiffuMT}\else \textit{URL withheld for double-blind review}\fi) as the trial interface.
The Space also serves as an open community challenge.
The analysis code that produces every number, figure and table in this section from the raw click log is released with the dataset.

\paragraph{Equivalence test.}
The headline binomial test in \cref{sec:twoafc} shows only the absence of a detectable difference from chance.
To obtain positive evidence of equivalence we run a two-one-sided test (TOST): aggregate accuracy is statistically equivalent to chance within a $\pm\afcEquivDelta{}$~percentage-point margin ($p=\afcTostP{}$), and the $90\%$ interval bounds the true accuracy to within $\pm\afcEquivMargin{}$~points of chance.
We fix the equivalence margin at $\pm\afcEquivDelta{}$~points (i.e.\ true accuracy below $\afcEquivHi{}$) \emph{a priori}; it lies well under the $64\%$ trainee discriminability that the comparable medical-imaging 2AFC study Med-cDiff~\citep{hung2023medcdiff} reports for a generator considered usable, so passing it is a conservative bar.
Per-expert accuracies ($\afcNTrials{}/\afcNExperts{}=10$ trials each) are individually uninformative and are reported (\cref{fig:2afc-per-expert}) only to show that the spread matches a binomial null, with individual accuracies ranging from \afcMinExpertAcc{} to \afcMaxExpertAcc{}.

\paragraph{Image pool.}
$66$ patches in total: \afcNReal{} real $256\!\times\!256$ patches drawn from the annotated test set used throughout the paper, and \afcNFake{} generated patches sampled from the diagnostic-selected checkpoint (offset~$0.2$, linear $\beta$-schedule, MSE loss, epoch~\runFifteenEpoch{}).
The generated patches are conditioned on the masks of the corresponding real patches, so the two pools share the underlying mask geometry.

\paragraph{Trial design.}
Each trial shows two patches side by side.
One is real, one generated, drawn from the respective pools and independently randomised left/right per trial.
Participants are asked ``\emph{Which of the two images is a real microtubule recording?}'' and click on one of the two patches.
No time limit is imposed; the median response time is \afcMedianRT{}~s.

\paragraph{Participants.}
\afcNExperts{} domain experts: a mix of cell-biology and microscopy researchers who work with \gls{irm} recordings of microtubule dynamics, and machine-learning researchers who are familiar with the output of diffusion models.
None has been involved in training the model or in selecting the checkpoint, and they were briefed only on the task, not on the appearance of the synthetic samples.
The paper authors are not part of the expert group.

\begin{figure}[tb]
\centering
\includegraphics[width=\linewidth]{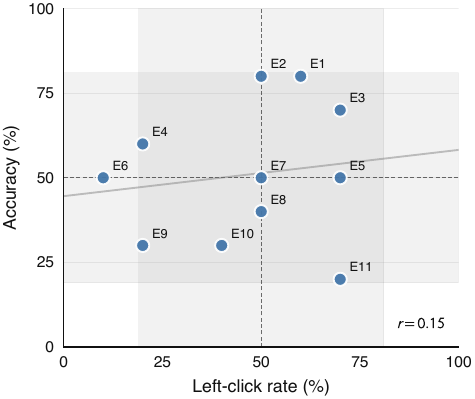}
\caption{\textbf{Chance-level accuracy is not an artifact of positional bias: per-expert accuracy and per-expert left-click rate are uncorrelated ($r=\afcLeftAccR{}$).}
Points cluster near the central crosshair $(50\%, 50\%)$ (chance accuracy and balanced clicking) with no systematic side preference.}
\label{fig:2afc-positional}
\end{figure}

\paragraph{Positional bias.}
We log the side of every click.
Across all trials, experts click the left slot \afcLeftRate{} of the time, and per-expert left-click rate is uncorrelated with accuracy ($r=\afcLeftAccR{}$, \cref{fig:2afc-positional}).
A constant left-clicking strategy would yield $50.0\%$ expected accuracy on a balanced pool, so the observed result is not an artifact of positional preference.

\paragraph{Engagement.}
The median reaction time is \afcMedianRT{}\,s (\afcMedianRTcorrect{}\,s on correct vs.\ \afcMedianRTincorrect{}\,s on incorrect trials), which indicates that experts deliberate rather than click through.

\paragraph{Anonymous-pool robustness.}
Additionally pooling the \afcNAnon{} anonymous trials logged by players who left before entering a name leaves the conclusion unchanged: \afcPooledAcc{} over \afcPooledN{} trials ($95\%$ CI $[\afcPooledCIlo{},\afcPooledCIhi{}]$, $p=\afcPooledBinomP{}$).

\section{Additional results}
\label{app:additional}

\paragraph{TSTR per-fold values.}
The medians reported in \cref{sec:tstr} summarize four recording-grouped folds; the individual values are real $\{0.80,0.60,0.51,0.72\}$, \synmtv\ $\{0.74,0.61,0.69,0.65\}$, real$+$\synmtv\ $\{0.78,0.67,0.69,0.68\}$, \synmtone\ $\{0.57,0.26,0.36,0.50\}$, and the Naive arm $\{0.61,0.36,0.49,0.55\}$.

\begin{figure}[tbp]
\centering
\includegraphics[width=\linewidth]{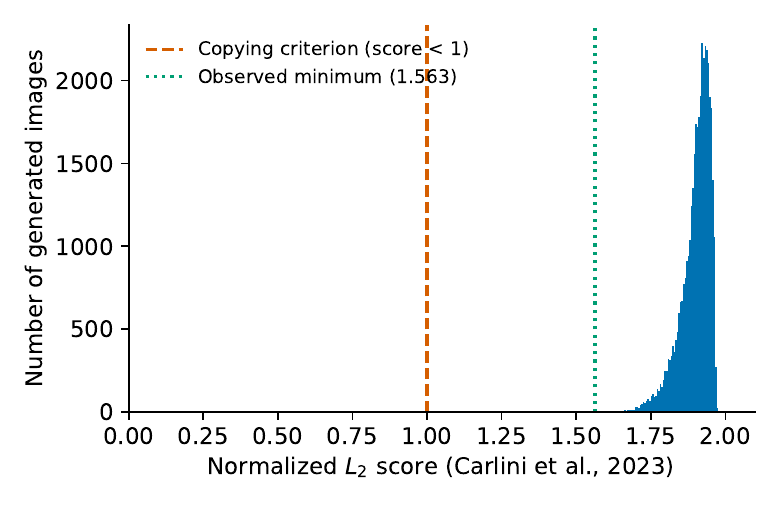}
\caption{\textbf{Every one of the $40{,}000$ generated images scores well above the copying threshold.} Distribution of the normalized $L_2$ memorization score~\citep{carlini2023extracting} across all generated images and checkpoints; scores below $1$ (dashed) would flag a potential copy. The observed minimum ($1.563$, dotted) is still far from that threshold.}
\label{fig:memorization-scores}
\end{figure}

\paragraph{Pixel-level memorization screen.}
The embedding-based memorization $\Delta$ above compares nearest-neighbor cosine similarities and cannot rule out near-exact pixel copies with dissimilar embeddings. We additionally screened for pixel-level memorization using the normalized $L_2$ nearest-neighbor criterion of \citet{carlini2023extracting} ($50$ nearest training neighbors, $\alpha=0.5$; a score below $1$ flags a potential copy), comparing $40{,}000$ generated images (all $200$ checkpoints) against the $278$ available training crops. No image met the copying criterion (minimum score $1.563$), while all ten exact-copy positive controls we inserted as a check were correctly flagged; \cref{fig:memorization-scores} shows the full score distribution.

\begin{figure*}[tb]
\centering
\includegraphics[width=\linewidth]{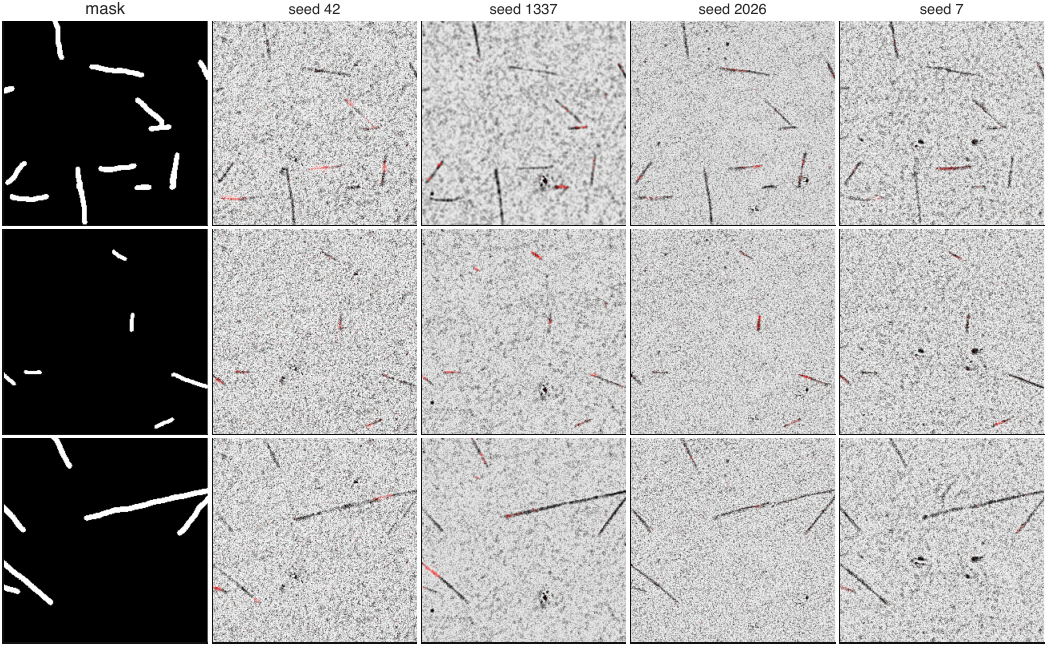}
\caption{\textbf{Same mask, different seeds (columns) yield visibly different textures and backgrounds (the diversity intra-sim measures); same seed, different masks (rows) share a diffuse background pattern because the seed fixes the initial latent $x_T$.}
Three masks (rows) $\times$ four seeds (columns), $50$ DDIM steps from the diagnostic-selected checkpoint (epoch~\runFifteenEpoch{}).}
\label{fig:diversity-grid}
\end{figure*}

\paragraph{Same mask, different seeds.}
The synthetic counterpart in each released triplet is one sample from $p(x \mid m)$, not a deterministic function of the mask.
\Cref{fig:diversity-grid} samples three masks at four independent seeds each; filament geometry is preserved across seeds while texture, background brightness, contrast and exact filament thickness vary.
The intra-set diversity flagged by the intra-similarity axis of our diagnostic (\cref{sec:diag}) is exactly this seed-to-seed variation aggregated over the test set.

Across all \gls{dm} runs of \cref{tab:absolute-runs}, the memorization $\Delta$~\citep{carlini2023extracting} (the nearest-neighbor cosine-similarity gap between train and test embeddings) stays at $0.01$, which rules out direct replication of training images.
Instance \gls{iou} ($0.32$--$0.37$) and the average per-image filament count ($\sim\!4$) likewise track the real reference (\gls{iou}~$0.37$, $4.10$ filaments per image).

\paragraph{Dataset composition.}
Each released \synmtv triplet pairs a binary mask with its real microscopy patch and one mask-conditioned synthetic sample (\cref{sec:model}).
A key advantage of pixel-space mask conditioning is that the release is not limited to a fixed tiling of the annotated data: because every labeled $512\times512$ crop is larger than the $256\times256$ model resolution, we sample \emph{arbitrarily many} random $256\times256$ views and synthesize a fresh conditional counterpart for each, turning $309$ hand-annotated crops into a much larger triplet corpus at no additional labeling cost.
Views are drawn only where the mask contains foreground (so every triplet shows filaments), and the $75$ source recordings are partitioned into disjoint train/validation/test groups \emph{before} cropping, so no recording---and therefore no underlying filament geometry---appears in more than one split.
The public release contains $2000$ training, $400$ validation and $400$ test triplets built this way; each synthetic image is generated with \gls{ddim} sampling from the diagnostic-selected checkpoint (epoch~\runFifteenEpoch{}), and filenames are anonymized to a running index so no acquisition metadata leaks.

\paragraph{Reproducibility.}
Random seeds were not fixed across the stochastic components (initial U-Net weights, forward noise, the initial sampling noise $x_T$; the \gls{ddim} reverse process itself is deterministic), though reading the diagnostic over $200$ samples per checkpoint keeps the per-axis numbers stable to this, and every trend reported in the paper holds across all eight hyperparameter configurations (\cref{sec:experiments}).
On publication we will release configuration files that fix the seeds and reproduce the numbers in \cref{tab:absolute-runs}, together with the diagnostic implementation, the 2AFC analysis code, and the trained checkpoint.

\end{document}